\documentclass[lettersize,journal]{IEEEtran}

\usepackage[dvips]{graphicx}
\usepackage[cmex10]{amsmath}
\usepackage{upgreek}
\usepackage{amsthm}
\usepackage{booktabs}
\usepackage{epsfig}
\usepackage{latexsym}
\usepackage{multirow}
\usepackage{stfloats}
\usepackage{epstopdf}
\usepackage{color}  
\usepackage{tabularx} 
\usepackage{amssymb}
\usepackage{enumerate}
\usepackage{enumitem}
\graphicspath{{./Figures/}}
\usepackage{color}
\usepackage{bbm}
\usepackage{bm}
\usepackage{cite}
\usepackage[tight,footnotesize]{subfigure}
\usepackage{balance}
\usepackage{mathrsfs}
\usepackage{verbatim}
\usepackage{dsfont}
\usepackage{verbatim}
\usepackage{tikz}
\usepackage{diagbox}
\usepackage{caption}
\allowdisplaybreaks[4]
\usepackage[framemethod=tikz]{mdframed}
\usepackage{multicol}
\usepackage{environ}
\usepackage{tikz}
\usepackage{algorithm}
\usepackage{algpseudocode}

\usepackage{subfig}
\usepackage{subcaption} 
\begin{document}

\title{Topology-Aware Two-Stage Federated Learning \\ via Proxy Models for Sub-THz \\ Heterogeneous LEO Communications}

\author{Jinhao Yi,~\IEEEmembership{Graduate Student Member,~IEEE}, Weijun Gao,~\IEEEmembership{Member,~IEEE}, \\ Chong Han,~\IEEEmembership{Senior~Member,~IEEE}, 
Ozgur Gurbuz,~\IEEEmembership{Senior~Member,~IEEE},
Josep M. Jornet,~\IEEEmembership{Fellow,~IEEE}
\thanks{
Jinhao~Yi and Weijun~Gao are with the Terahertz Wireless Communications (TWC) Laboratory, Shanghai Jiao Tong University, Shanghai 200240 China (email:~\{jinhao.yi, gaoweijun\}@sjtu.edu.cn). 

Chong~Han is with the Terahertz Wireless Communications (TWC) Laboratory, Cooperative Medianet Innovation Center (CMIC), School of Information Science and Electronic Engineering, Shanghai Jiao Tong University, Shanghai 200240 China (email:~chong.han@sjtu.edu.cn). 

Ozgur Gurbuz is with
the Electronics Engineering Program, Faculty of Engineering and Natural
Sciences, Sabanci University, Istanbul 34956, Türkiye (e-mail:~ogurbuz@sabanciuniv.edu).

Josep M. Jornet is with the Institute for the Wireless Internet of Things, Northeastern University, Boston, MA 02115 USA (email:~j.jornet@northeastern.edu). 
}
}

\maketitle

\begin{abstract}
Federated learning (FL) has emerged as a promising distributed training paradigm for Low Earth Orbit (LEO) networks by significantly reducing communication overhead. However, its deployment faces critical challenges, e.g., topology-induced model staleness, short contact windows, and unaddressed computing heterogeneity. To address these issues, a topology-aware two-stage FL framework is proposed in this paper. First, a multi-layer physical architecture utilizing high-altitude platforms (HAPs) and Sub-THz communications is designed to extend satellite-ground contact windows and enlarge available bandwidth. Second, a proxy-model-based approach is adopted to fully utilize heterogeneous resources and enable architecture-agnostic knowledge aggregation.
Finally, building upon these foundations, a topology-aware two-stage aggregation mechanism is proposed as the central algorithmic design to overcome the topology-induced staleness. The mechanism dynamically partitions LEO satellites into localized groups based on their transient HAP coverage. Within each group, LEO satellites perform asynchronous aggregation at their associated HAP to naturally tolerate computational delays without penalizing faster nodes. Subsequently, a synchronous inter-group aggregation is executed among all HAPs at the Ground Station (GS) to strictly bound the maximum staleness and guarantee stable global convergence. 
Numerical results demonstrate the proposed framework extends contact windows and achieves 86.59\%--90.57\% test accuracy, outperforming the state-of-the-art heterogeneous baseline by 16.26\%--19.80\%. Furthermore, it achieves a 1.5$\times$ to 2.2$\times$ convergence speedup, which closely approaches the ideal upper bound.
\end{abstract}


\section{Introduction}
\IEEEPARstart{W}{ith} the rapid development of next-generation communication technologies~\cite{yang20196g,dang2020should}, satellite technology~\cite{fang2022olive,zhai2023fedleo} has been widely recognized as a key enabler for achieving seamless connectivity beyond terrestrial networks~\cite{zhao2019uav,zhu2020millimeter}. Recent years have witnessed the explosive deployment of low Earth orbit (LEO) satellites by commercial companies (e.g., SpaceX, OneWeb) and governmental organizations (e.g., ESA, NASA) to provide high-reliability communication and global coverage~\cite{ahmmed2022digital}. Equipped with advanced sensors and high-resolution cameras, LEO satellites can continuously collect massive volumes of Earth imagery and environmental data with high-resolution~\cite{chen2022robust, wu2024accelerating, yuan2023graph}.
To support the explosive growth of space-generated data and high-throughput applications, emerging constellations are increasingly adopting sub-Terahertz (sub-THz) bands~\cite{11242136,gao2025terahertz,fcc2025spectrum}, which provide ultra-wide contiguous spectrums to overcome the severe bandwidth scarcity of traditional Ku/Ka bands.
Empowered by these rich datasets and high-capacity THz links, a wide range of applications, including environmental monitoring, disaster response, and climate analysis,  can be profoundly advanced utilizing the power of deep learning~\cite{liu2021uav,liu2021novel,huang2024energy}.

Despite the integration of high-bandwidth links, the unprecedented surge in data generated by LEO satellites also brings critical new challenges.~\cite{letaief2021edge,chen2021rf}. 
Specifically, transmitting massive raw data to a ground station (GS) easily overwhelms downlink capacities and introduces unacceptable latency~\cite{so2022fedspace,vasisht2021l2d2}.
To overcome this fundamental bottleneck, federated learning (FL) has emerged as a promising distributed training paradigm to address these challenges~\cite{mcmahan2017communication,chen2020joint,xu2020client}. In FL, each satellite trains a local model using its own data and only transmits model parameters to a GS for aggregation, which significantly reduces communication overhead compared with centralized training.  

Nevertheless, applying traditional federated learning frameworks, such as FedAvg~\cite{mcmahan2017communication} and FedAsync~\cite{xie2019asynchronous}, directly to LEO satellite networks can still face several critical challenges. First, communication constraints remain a primary hurdle. Despite the wide bandwidth of sub-THz links, the highly transient contact windows of LEO satellites still bottleneck multi-round model transmissions.
Second, satellites launched in different phases possess vastly different computational capacities, memory limits, and energy budgets. The extreme heterogeneity among diverse satellites renders traditional homogeneous FL frameworks inapplicable. 
Third, and most critically, the highly dynamic 3D orbital topology severely exacerbates model staleness~\cite{li2020federated}. Beyond merely waiting for computing stragglers, satellites suffer from asymmetric visibility and intermittent link disconnections. As a result, locally trained updates are frequently trapped in prolonged communication blind zones, rendering them obsolete and degrading global aggregation. To address these challenges, existing studies on LEO-specific FL have primarily focused on alleviating communication constraints via advanced physical-layer techniques (e.g., inter-satellite links~\cite{zhai2023fedleo}, NOMA-OFDM~\cite{elmahallawy2024communication}) or mitigating staleness through simple time-dependent weighting functions~\cite{elmahallawy2022asyncfleo,razmi2022scheduling}. However, these frameworks predominantly assume homogeneous satellite architectures. While a few works~\cite{lin2024fedsn} explore model heterogeneity via sub-structure extraction, they struggle with complex architectures like Transformers. Conversely, terrestrial proxy-model paradigms~\cite{xie2024mh,li2025mergenet} offer powerful architecture-agnostic solutions. However, naively migrating them to LEO networks inevitably triggers severe topology-induced staleness due to intermittent space links. 

To bridge these gaps, we propose a topology-aware two-stage federated learning framework for Sub-THz heterogeneous LEO networks to jointly address communication bottlenecks, hardware heterogeneity, and model staleness.
First, as a physical foundation, we design a satellite-HAP-GS network that utilizes HAP relays and Sub-THz links to extend contact windows and enlarge bandwidth~\cite{fcc2025spectrum}.
Second,
to explicitly address the heterogeneity of LEO satellites, we adapt a proxy-model-based FL paradigm to enable architecture-agnostic knowledge aggregation, while efficiently exploiting the different onboard resources. 
Crucially, building upon the physical foundation and the adapted proxy mechanism, we develop a topology-aware two-stage aggregation mechanism. By dynamically partitioning satellites based on HAPs, this mechanism performs asynchronous intra-group aggregation to tolerate delays, followed by synchronous inter-group aggregation to strictly bound staleness. This hierarchical design effectively guarantees stable global convergence with high accuracy. Specifically, the contributions of this paper are summarized as follows.

\begin{itemize}
\item \textbf{We propose a sub-THz satellite-HAP-GS integrated network as the physical foundation for satellite FL.} By leveraging HAP relays and sub-THz links, this multi-layer architecture significantly extends contact windows and boosts capacity, fundamentally alleviating transmission bottlenecks and establishing a crucial basis for topology-aware aggregation mechanism.

\item \textbf{We adapt a proxy-model-based FL paradigm to overcome extreme hardware heterogeneity across satellites.} Rather than enforcing uniform local models or rigid sub-structure partitioning, this paradigm facilitates architecture-agnostic knowledge exchange and allows satellites with vastly different computational capacities to participate efficiently.

\item \textbf{We develop a topology-aware two-stage aggregation mechanism to jointly mitigate model staleness and preserve aggregation accuracy in dynamic LEO satellite networks.}
By partitioning satellites into topology-aware groups according to their associated HAP relays, the mechanism first performs asynchronous aggregation within each group to improve efficiency, followed by synchronous aggregation across different groups to preserve global model accuracy. This hierarchical design effectively leverages the network topology to accommodate heterogeneous contact opportunities and dynamic satellite connectivity, achieving a favorable balance between efficiency and aggregation precision.

\item \textbf{We evaluate extensive numerical results to validate the proposed heterogeneous FL framework over Sub-THz SAGIN.} Based on the Walker constellation model and the public EuroSAT remote sensing dataset, experimental results demonstrate that the proposed framework achieves a superior average test accuracy of 90.57\%, outperforming the state-of-the-art heterogeneous baseline by up to 19.80\%. Furthermore, the two-stage mechanism accelerates convergence speed by over 2.2$\times$ while maintaining a tight accuracy consistency margin of 7.05\% across drastically heterogeneous satellites.

\end{itemize}



The remainder of the paper is organized as follows. Sec.~\ref{system_model} introduces the system model. Sec.~\ref{proxy_model} introduces the adopted proxy-model-based FL framework, and Sec.~\ref{aggregation} details the two-stage, topology-aware aggregation mechanism. Sec.~\ref{numerical} presents the numerical results, followed by conclusions in Sec.~\ref{conclusion}.

\begin{figure*}[t]
\centerline{\includegraphics[width=0.7\textwidth]{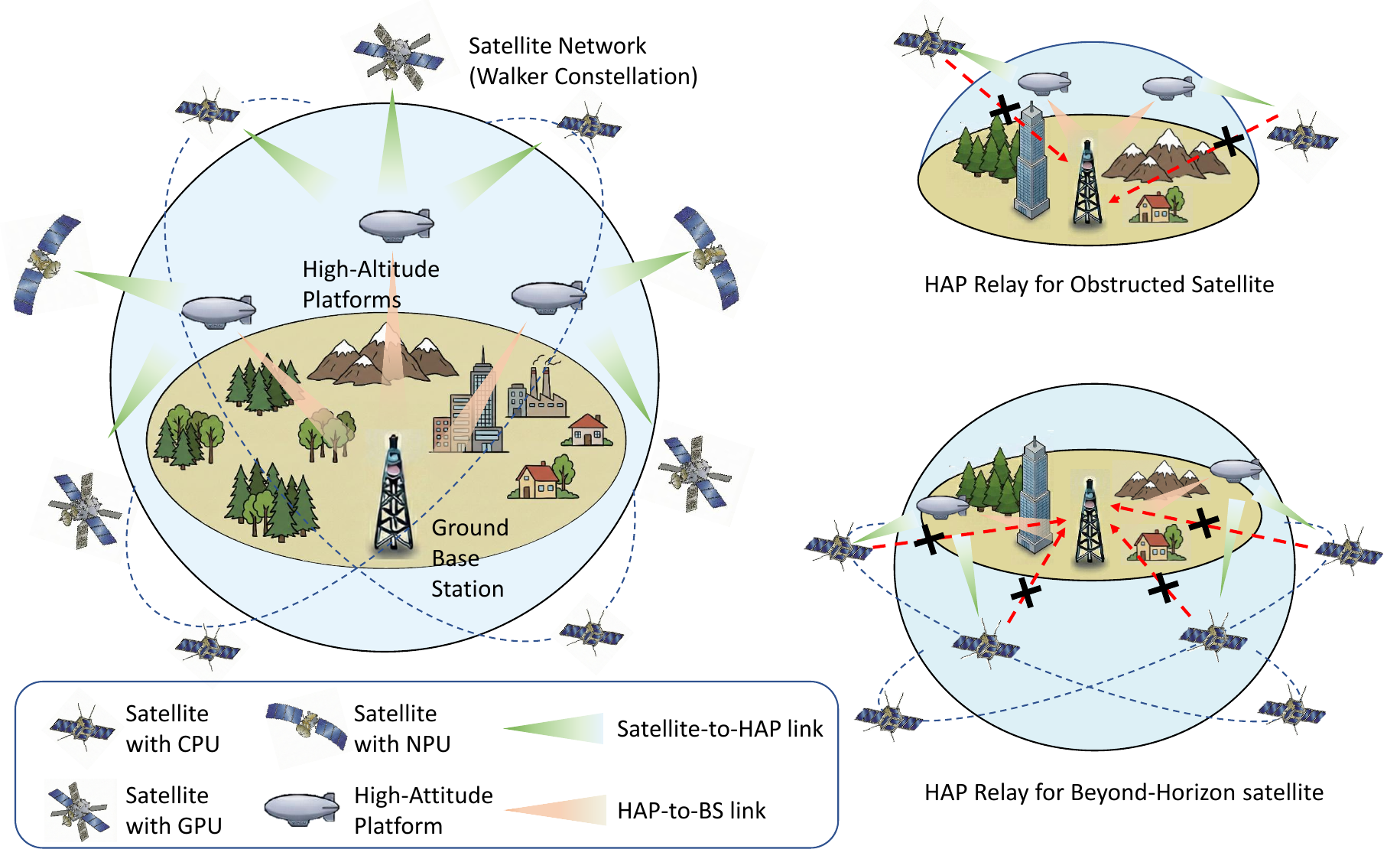}}
\captionsetup{font={footnotesize}}
\caption{System model of the satellite-HAP-GS integrated network, where HAPs act as intermediate relays to ensure continuous global connectivity, specifically extending communication windows for obstructed satellites and beyond-horizon satellites.}
\label{fig:system_model}
\end{figure*}

\section{System Model}
\label{system_model}
As illustrated in Fig.~\ref{fig:system_model}, we consider a multi-altitude 3D Walker-like LEO satellite network with many HAPs and a GS. Satellites perform local model training using onboard data, while GS serves as the global aggregation center. To overcome the limited satellite-GS visibility and enhance link capacity, HAPs are deployed to act as intermediate relays, and Sub-THz communications are employed to support efficient model transmission and aggregation. A detailed description of the system model is presented as follows.

\subsection{Satellite-HAP-GS Multi-Layer Model}
In our scenario, we consider the satellite-HAP-GS network comprising $N_s$ LEO satellites and $N_h$ HAPs serving as intermediate relays to a GS. Specifically, the constellation of $N_s$ LEO satellites is deployed according to a multi-altitude Walker-like configuration~\cite{wei2020efficient,mo2016multi}, where satellites are assumed to be standard CubeSats to align with practical deployment scenarios, and are distributed uniformly across $P$ orbital planes at different altitudes $H_s\in\{h_1,h_2,\dots,h_n\}$.
Hence, the right ascension of the ascending node (RAAN) $\Omega^m_i$ and the mean anomaly $u^m_i$ of the $i$-th satellite in the $m$-th plane are governed by
\begin{equation} 
\left\{\begin{array}{l} 
\Omega^m_i=\Omega^m_{0}+(m-1) \frac{2 \pi}{P} \\ 
u^m_i=u^m_{0}+(m-1) F \frac{2 \pi}{T}+(i-1) P \frac{2 \pi}{T} 
\end{array}\right. 
\end{equation}
where $\Omega_{0}$ and $u_{0}$ denote the reference RAAN and initial mean anomaly, respectively. $T=N_s/P$ denotes the number of satellites in each plane and $F$ represents the inter-plane phasing factor.

Building upon this foundational spatial topology, the state of the $i$-th satellite in the $m$-th plane over time can be characterized by its orbital parameter tuple $\boldsymbol{\Theta}^m_i(t)$. It consists of the semi-major axis $a_m$, the RAAN $\Omega^m_i$, and the time-varying mean anomaly $u^m_i(t)$, and is given as 
\begin{equation}
\begin{aligned}
\boldsymbol{\Theta}^m_i(t) &\triangleq \left\{ a_m, \Omega^m_i, u^m_i(t) \right\} \\
&= \Bigg\{ R_\text{Earth} + H_s^m, \Omega^m_0 + (m-1)\frac{2\pi}{P}, \\
&\qquad u_0 + \frac{2\pi}{T}  \big[(m-1)F + (i-1)P\big] + \\
&\qquad \sqrt{\frac{\mu}{(R_{Earth} + H_s^i)^3}}t \Bigg\}
\end{aligned}
\end{equation}
where $H_s^m \in \{h_1, \dots, h_n\}$ denotes the specific altitude of the $m$-th plane, $R_\text{Earth} = 6371$~km is the Earth's radius, and $\mu = 398600.4418~\text{km}^3/\text{s}^2$ is the Earth's gravitational parameter. This unified tensor representation preserves rigorous orbital tracking while maintaining tractable visibility analysis. Furthermore, we consider the $N_h$ quasi-stationary HAPs deployed at a fixed altitude $H_h$ within the line-of-sight~(LoS) of GS. Acting as intermediate relays, these HAPs connect local training satellites to the GS, enabling efficient federated learning across the integrated architecture.

\subsection{Sub-THz Channel Model}
To facilitate high-rate model transmission, we adopt Sub-THz communications, particularly the W-band for both satellite-HAP links. The W-band (spaning from 94.1-100~GHz) provides substantially higher data rates than traditional Ku/Ka bands due to its ultra-wide effective bandwidth~$\mathcal{W}$ of 5.9~GHz, as proposed by the Federal Communications Commission~(FCC)~\cite{fcc2025spectrum}. Besides, it also exhibits superior resilience to atmospheric disturbances and spatial interference compared to optical wireless communications. This unique combination of ultra-high capacity and robust link reliability enables fast and secure model transmission and aggregation for federated learning in dynamic space environments.

According to~\cite{kokkoniemi2021channel,Masihi2025Terahertz}, The channel capacity is fundamentally constrained by severe free-space path loss, molecular absorption, and beam misalignment. Hence, the instantaneous Shannon capacity $\mathcal{C}_{i,n}(t)$ between the $i$-th satellite and the $n$-th HAP over~$\mathcal{W}$ is given as
\begin{equation}
\begin{aligned}
\mathcal{C}_{i,n}(t) &= \int_{\mathcal{W}} \log_2 \Bigg( 1 + \frac{\mathcal{P}_{tx}(f) G_{Tx}(\theta_{i,n}^{tx}) G_{Rx}(\theta_{i,n}^{rx}) c^2}{\left(4\pi f \|\mathbf{r}_i(t) - \mathbf{r}_n\| \right)^2} \\
&\qquad \times \frac{h_{pe}(\alpha(t))}{\Psi(f, \mathbf{r}_i, \mathbf{r}_n) \mathcal{N}_{tot}(f, T)} \Bigg) df
\end{aligned}
\label{capacity}
\end{equation}
where $\mathbf{r}_i(t)$ and $\mathbf{r}_n$ are the position vectors of the $i$-th satellite and the $n$-th HAP, respectively. $\mathcal{P}_{tx}(f)$ represents the transmit power spectral density of the satellite, and $T$ denotes the atmospheric temperature. $G_\text{Tx}(\theta_\text{Tx})$ and $G_\text{Rx}(\theta_\text{Rx})$ are the transmitter and receiver antenna gains, respectively. The term $\Psi(f, \mathbf{r}_i, \mathbf{r}_m)$ represents the cumulative multi-species molecular absorption along the propagation path, which is given as 
\begin{equation}
\Psi(f, \mathbf{r}_i, \mathbf{r}_n) = \exp\left( \int_{0}^{\|\mathbf{r}_i(t) - \mathbf{r}_n\|} \sum_{q} \kappa_a^{(q)}(f, \ell) d\ell \right),
\end{equation}
where $\kappa_a(f,r_{atm}) = \sum_i\kappa_a^i(f,r_{atm})$ is the total summed molecular absorption coefficient along the propagation path through the atmosphere. In addition, $h_{pe}(\alpha(t))$ represents the pointing loss factor due to beam misalignment. Given the narrow half-power at Sub-THz bandwidth $\mathcal{W}$, the pointing-induced gain loss is a first-order impairment, which is expressed as~\cite{Masihi2025Terahertz}
\begin{equation}
h_{pe}(\alpha(t)) = \exp\left(-2 \frac{(\|r_i(t)-r_n\| \tan \alpha(t))^2}{w_{zeq}^2}\right)
\end{equation}
where $\alpha(t)$ is the instantaneous pointing error angle, and $w_{zeq}^2$ is the equivalent beam waist given as
\begin{equation}
w_{zeq}^2 = \frac{w_z^2 \sqrt{\pi} \text{erf}(v)}{2v \exp(-v^2)}, \quad v = \sqrt{\frac{\pi}{2}} \left( \frac{a}{w_z} \right)
\end{equation}
where $a$ is the receiver aperture radius and $w_z \approx \frac{c z}{\pi f w_0}$ is the beam waist at distance $z$, with $w_0$ being the beam waist at the transmitter related to the antenna gain $G_{Tx}$. Besides, the term $\mathcal{N}_{tot}(f, T)$ represents the aggregated noise spectral density. Since it is negligible at Sub-THz bandwidth $\mathcal{W}$, the noise is dominated by classical thermal characteristics given as
\begin{equation}
\mathcal{N}_{tot}(T) = k_B T 10^{N_f/10}
\end{equation}

Furthermore, while atmospheric turbulence can induce signal scintillation, the scintillation index at W-band frequencies for satellite-to-HAP downlinks is on the order of $10^{-5}$~\cite{Aliaga2025Analysis}. Hence, this effect is practically negligible, and is therefore omitted from our channel model. Besides, it is essential to clarify that the HAP-GS link utilizes the same fundamental channel capacity formulation as~\eqref{capacity}, but experiences significantly more severe molecular absorption due to traversing the water-vapor-filled troposphere.

\subsection{Visibility and Contact Model}
For communication to be established between a LEO satellite and a HAP or GS, a LoS condition must be satisfied. In our model, we adopt an elevation-angle-based visibility model based on a 3D Earth-centered coordinate system to characterize satellite-HAP and satellite-GS connectivity. Fundamentally, a reliable LoS contact is established only when the relative elevation angle between the communicating entities exceeds a predefined minimum threshold. Let $\mathbf{r}_g$ denote the GS located at a fixed position on the Earth's surface. The elevation angle between the $i$-th satellite and the $n$-th HAP, denoted as $\theta_{i,n}(t)$, is defined as
\begin{equation}
\theta_{i,n}(t)
=
\arccos
\left(
\frac{
(\mathbf{r}_i(t)-\mathbf{r}_n) \cdot \mathbf{r}_n
}{
\|\mathbf{r}_i(t)-\mathbf{r}_n\| \, \|\mathbf{r}_n\|
}
\right)
- \frac{\pi}{2}.
\end{equation}
Consequently, the $i$-th satellite and the $n$-th HAP are considered to be in LoS contact at time $t$ if and only if the following condition is satisfied $\theta_{i,n}(t) \ge \theta_{\min}^{\mathrm{SH}}$, where $\theta_{\min}^{\mathrm{SH}}$ denotes the strictly required minimum elevation angle threshold to maintain a reliable space-to-air link. Similarly, the $i$-th satellite is considered visible to the GS if the elevation angle $\theta_{n,g}$ satisfies $\theta_{i,g} \ge \theta_{\min}^{\mathrm{HG}}$,
where $\theta_{\min}^{\mathrm{SG}}$ is the minimum elevation angle threshold for the satellite-GS link. For the HAP-GS link, it is worth noting that the HAPs are predefined to be deployed within the LoS coverage of the GS. Hence the visible condition of HAP-GS link is assumed to be always satisfied in the considered system model. Similarly, for the direct satellite-GS link (which serves as a conventional baseline configuration), its visibility indicator $\mathbb{I}_{i,g}(t)$ is exclusively activated when $\theta_{i,g}(t) \ge \theta_{\min}^{\mathrm{SG}}$.

Building upon these instantaneous LoS conditions, the overall connectivity state of the communication paths can be strictly determined. Let $\mathbb{I}_{i,n}(t) \in \{0, 1\}$ denote the binary indicator for the relayed link via the $n$-th HAP, and $\mathbb{I}_{i,g}(t) \in \{0, 1\}$ denote the indicator for the direct satellite-GS link. These are formulated as
\begin{equation}
\mathbb{I}_{i,n}(t) = 
\begin{cases} 
1, & \text{if } \theta_{i,n}(t) \ge \theta_{\min}^{\mathrm{SH}} \\
0, & \text{otherwise}
\end{cases}
\label{vis_1}
\end{equation}

\begin{equation}
\mathbb{I}_{i,g}(t) = 
\begin{cases} 
1, & \text{if } \theta_{i,g}(t) \ge \theta_{\min}^{\mathrm{SG}} \\
0, & \text{otherwise}
\end{cases}.
\label{vis_2}
\end{equation}
To fully exploit the integrated space-air-ground architecture, the $i$-th satellite can successfully communicate with the GS if either the direct link or the HAP-relayed link is available. Hence, the integrated effective contact window, denoted as $\mathcal{T}_{vis}^{i}$, can be obtained by embedding the spatial geometry of both links and logical union directly into the temporal integration, which is given as
\begin{equation}
\begin{aligned}
\mathcal{T}_{vis}^{i} &= \int_{0}^{\mathcal{T}_{orb}^i} \max \Bigg\{ \\
&\quad \mathbb{I} \Bigg( \arccos\left(\frac{(\mathbf{r}_i(t)-\mathbf{r}_n) \cdot \mathbf{r}_n}{\|\mathbf{r}_i(t)-\mathbf{r}_n\| \, \|\mathbf{r}_n\|}\right) - \frac{\pi}{2} \ge \theta_{\min}^{\mathrm{SH}} \Bigg), \\
&\quad \mathbb{I} \Bigg( \arccos\left(\frac{(\mathbf{r}_i(t)-\mathbf{r}_g) \cdot \mathbf{r}_g}{\|\mathbf{r}_i(t)-\mathbf{r}_g\| \, \|\mathbf{r}_g\|}\right) - \frac{\pi}{2} \ge \theta_{\min}^{\mathrm{SG}} \Bigg) \Bigg\} dt
\end{aligned}
\label{vis_all}
\end{equation}
where $\mathcal{T}_{orb}^i = 2\pi\sqrt{a_i^3/\mu}$ is the complete orbital period of the $i$-th satellite. This integral precisely captures the enhanced, heterogeneous contact durations experienced by different satellites under the satellite-HAP-GS integrated
network. 

With the multi-layer physical architecture established and the transient visibility windows $\mathcal{T}_{vis}^{i}$ rigorously derived, these spatial constraints directly motivate our topology-dependent heterogeneity formulation and the subsequent design of our adapted proxy-model-based FL paradigm.

\begin{figure}[t]

\centerline{\includegraphics[width=0.5\textwidth]{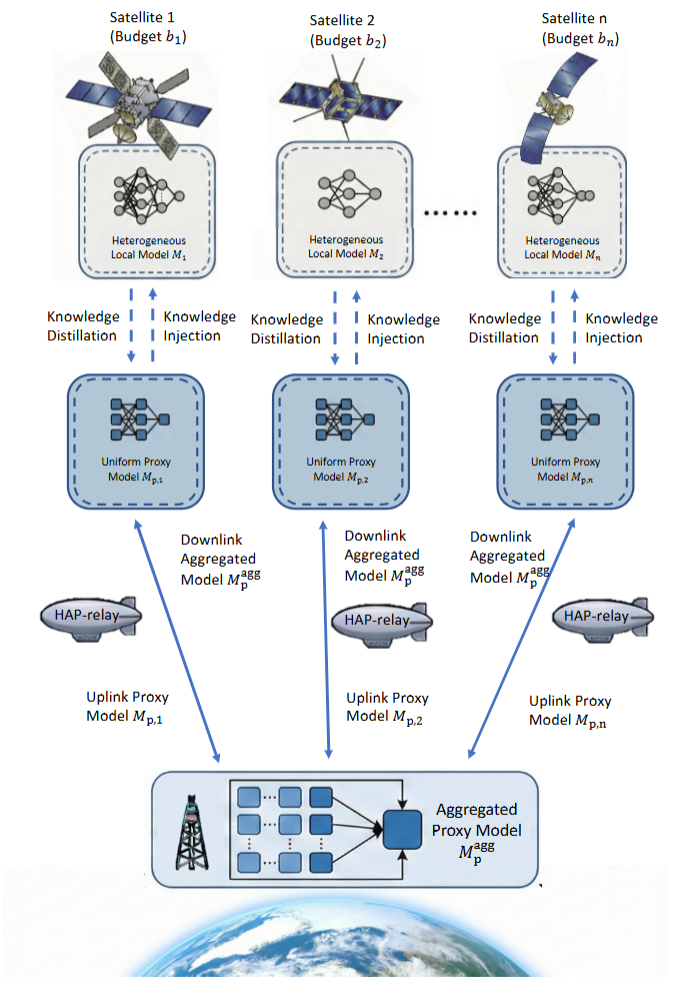}}

\captionsetup{font={footnotesize}}

\caption{Overview of the proxy-assisted FL framework enabling effective knowledge transfer across heterogeneous satellites.}

\label{fig:fed}

\end{figure}
\section{Proxy-Model-Based Federated Learning for Heterogeneous Satellites}
\label{proxy_model}
\begin{figure}[t] 

\centering

\subfigure[]{

\includegraphics[width=1\columnwidth]{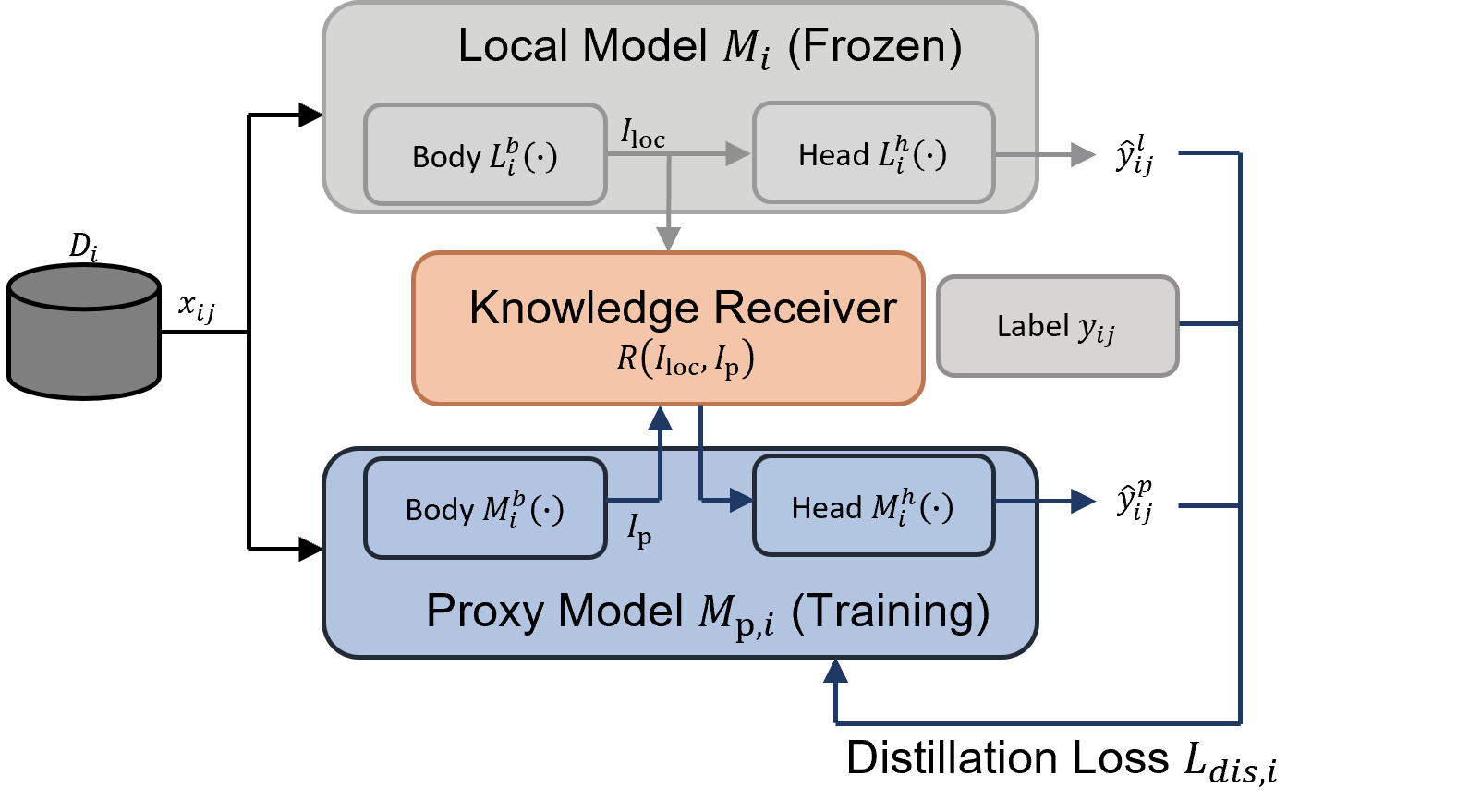} 

}
\\[15pt] 
\subfigure[]{

\includegraphics[width=1\columnwidth]{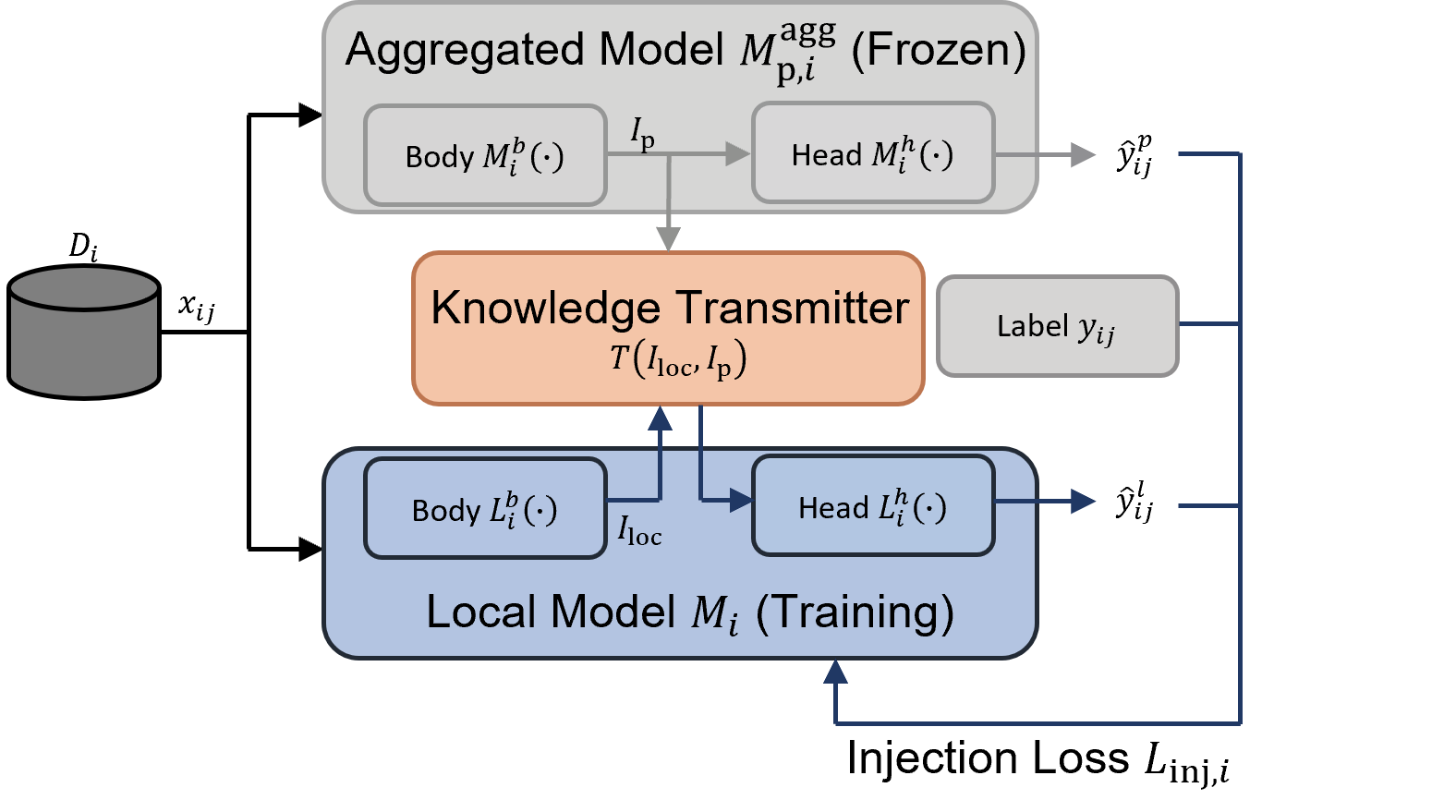} 

}

\captionsetup{font={footnotesize}}

\caption{Detailed process of knowledge transfer. (a) knowledge distillation from local model to proxy model; (b) knowledge injection from proxy model to local model.}

\label{fig_knowledge}


\end{figure}

Driven by the physical foundation established in Sec.~\ref{system_model}, 
we first quantitatively formulate the hardware and topology-dependent heterogeneity of LEO satellites. To overcome these architectural gaps, we further adapt a proxy-model-based FL paradigm inspired by recent terrestrial FL study~\cite{xie2024mh,li2025mergenet}. 
As depicted in Fig.~\ref{fig:fed} and Fig.~\ref{fig_knowledge}, this lightweight paradigm facilitates architecture-agnostic knowledge exchange, enabling the efficient participation of these heterogeneous satellites and establishing the foundation for our topology-aware aggregation strategy. The complete architecture and interaction processes of the adapted framework are detailed as follows.

\subsection{Heterogeneity in LEO Satellites}
In the considered large-scale LEO satellite network, heterogeneity naturally arises from both onboard hardware constraints (i.e., computation and memory) and dynamic 3D orbital topologies. Satellites launched in different phases possess varying processors (e.g., CPUs to NPUs) and memory limits, causing disparate local training speeds. Concurrently, diverse orbital planes yield highly varying visibility windows with HAPs, dictating uneven communication opportunities and model update frequencies. To characterize these heterogeneous constraints quantitatively, let $\boldsymbol{\rho}_i \triangleq [\rho_i^{\text{cmp}}, \rho_i^{\text{mem}}, \rho_i^{\text{com}}]^\top \in [0,1]^3$ denote the normalized computing, memory, and communication capability vector of the $i$-th satellite. While foundational studies like~\cite{lin2024fedsn} rely on static parameters, we elevate this paradigm by constructing a \emph{spatio-temporal coupled capability space}. By rigorously embedding the transient visibility window $\mathcal{T}_{vis}^{i}$ and instantaneous Sub-THz capacity $\mathcal{C}_{i,n}(t)$ derived in Sec.~\ref{system_model}, the topology-dependent capabilities are unified as:
\begin{equation}
\left\{
\begin{aligned}
\rho_i^{\text{cmp}} &= \min \left\{ 1, \frac{1}{\xi \vartheta} \int_{\mathcal{T}_{vis}^{i}} \omega_i^{\text{cmp}}(t) dt \right\} \\
\rho_i^{\text{mem}} &= \min \left\{ 1, \frac{\Omega_i}{\Phi_{\text{glb}}} \right\} \\
\rho_i^{\text{com}} &= \min \left\{ 1, \frac{1}{\Phi_{\text{loc}}} \int_{\mathcal{T}_{vis}^{i}} \mathcal{C}_{i,n}(t) dt \right\}
\end{aligned}
\right., 
\end{equation}
where $\omega_i^{\text{cmp}}(t)$ is the instantaneous processing capability (FLOPs/s) allocated for local training, $\xi$ and $\vartheta$ denote the local epochs and computational workload per mini-batch. $\Omega_i$ and $\Phi_{\text{glb}}$ ($\Phi_{\text{loc}}$) represent the available onboard memory and the specific model data sizes, respectively. 

Ultimately, dictated by the stringent intersection of physical topology and hardware limits, the effective dynamic training budget $\mathcal{B}_i$ is strictly bounded by the minimum envelope of this capability vector:
\begin{equation}
\mathcal{B}_i = \min \left\{ \rho_i^{\text{cmp}}, \rho_i^{\text{mem}}, \rho_i^{\text{com}} \right\}, \quad \forall i \in \{1,\dots,N_s\}.
\end{equation}

This cross-layer formulation compactly bridges 3D orbital dynamics with localized hardware disparities, establishing the rigorous quantitative foundation for our adapted proxy-model framework.

\subsection{Adapted Proxy-Model-Based Knowledge Exchange}
Guided by the dynamic budget~$\mathcal{B}_i$, we implement an adapted proxy-model-based mechanism to facilitate architecture-agnostic knowledge exchange. Rather than enforcing uniform local models, we design local models of different sizes $M_i$ for each satellite $i$ according to their available budget~$\mathcal{B}_i$, alongside a lightweight, uniform proxy model $M_{\mathrm{p},i}$ sized to satisfy the minimal communication budget $\mathcal{B}_\text{min}$ among satellites. More importantly, the proxy model decouples knowledge exchange into two stages: \emph{knowledge distillation}, where local models transfer learned features to the proxy for global aggregation; and \emph{knowledge injection}, where the globally aggregated proxy guides local training. By transmitting only these lightweight proxy models, our method significantly reduces communication overhead and enables architecture-agnostic aggregation. We detail these two stages as follows.

\subsubsection{Knowledge Distillation from Local Model to Proxy Model}
During the \emph{knowledge distillation} stage, the local model $M_i$ on each LEO satellite $i$ is first trained using its local dataset $\mathcal{D}_i = \{(x_{ij}, y_{ij})\}_{j=1}^{|\mathcal{D}_i|}$. To transfer the learned knowledge from the local model $M_i$ to its corresponding proxy model $M_{\mathrm{p},i}$, we freeze the parameters of $M_i$ to train $M_{\mathrm{p},i}$. To capture the inter-class relational knowledge, we define the softened probability distribution incorporating a temperature scaling factor $\tau$ as
\begin{equation}
q_{ij,c}(\boldsymbol{z}_{ij},\tau) = \frac{\exp(z_{ij,c} / \tau)}{\sum_{k=1}^C \exp(z_{ij,k} / \tau)},
\end{equation}
where $C$ is the total number of classes, $\boldsymbol{z}_{ij} = [z_{ij,1}, \dots, z_{ij,C}]^\top$ denote the logit output vector, and $z_{ij,c}$ represents the logit output for the $c$-th class, which serves as the core medium for architecture-agnostic knowledge exchange. Consequently, the distillation loss $\mathcal{L}_{\mathrm{dis},i}$ for satellite $i$ is formulated by combining the standard Cross-Entropy (CE) loss and the Kullback-Leibler (KL) divergence, defined as
\begin{equation}
\begin{split}
\mathcal{L}_{\mathrm{dis},i} &= -\lambda^p_{\mathrm{dis}} \sum_{j=1}^{|\mathcal{D}_i|} \sum_{c=1}^C y_{ij,c} \log p^p_{ij,c}(\boldsymbol{z}^p_{ij}) \\
&+ \lambda^{\mathrm{con}}_{\mathrm{dis}} \tau^2 \sum_{j=1}^{|\mathcal{D}_i|} \sum_{c=1}^C q^l_{ij,c}(\boldsymbol{z}_{ij},\tau) \log \left( \frac{q^l_{ij,c}(\boldsymbol{z}^l_{ij},\tau)}{q^p_{ij,c}(\boldsymbol{z}^p_{ij},\tau)} \right),
\end{split}
\end{equation}
where $p^p_{ij,c}$ is the standard softmax output ($\tau=1$) of the proxy model, $q^l_{ij,c}(\tau)$ and $q^p_{ij,c}(\tau)$ are the softened probabilities of the frozen local model and the proxy model, respectively, and $\lambda^p_{\mathrm{dis}}, \lambda^{\mathrm{con}}_{\mathrm{dis}}$ are weighting factors balancing the two objectives.

\subsubsection{Knowledge Injection from Proxy Model to Local Model}
During the \emph{knowledge injection} stage, each satellite receives the globally aggregated proxy model, denoted as $M_{\mathrm{gp}}$, from the GS. This global proxy model encapsulates the collective distilled knowledge from all participating satellites. To inject this structural knowledge back into each local model $M_i$, we freeze $M_{\mathrm{gp}}$ and use it to supervise the local training process on $\mathcal{D}_i$. The injection loss $\mathcal{L}_{\mathrm{inj},i}$ is formulated to enforce consistency between the local predictions and the global guidance, defined as
\begin{equation}
\begin{split}
\mathcal{L}_{\mathrm{inj},i} &= -\alpha \sum_{j=1}^{|\mathcal{D}_i|} \sum_{c=1}^C y_{ij,c} \log p^l_{ij,c}(\boldsymbol{z}_{ij}) \\
& + (1 - \alpha) \tau^2 \sum_{j=1}^{|\mathcal{D}_i|} \sum_{c=1}^C q^{\mathrm{gp}}_{ij,c}(\boldsymbol{z}^{\mathrm{gp}}_{ij},\tau) \log \left( \frac{q^{\mathrm{gp}}_{ij,c}(\boldsymbol{z}^{\mathrm{gp}}_{ij},\tau)}{q^l_{ij,c}(\boldsymbol{z}^l_{ij},\tau))} \right),
\end{split}
\end{equation}
where $p^l_{ij,c}$ represents the standard prediction of the local model, $q^{\mathrm{gp}}_{ij,c}(\tau)$ denotes the soft targets generated by the aggregated global proxy model, and $\alpha \in (0,1)$ is the injection coefficient that balances the empirical risk minimization and the global knowledge assimilation.  

\subsection{Architecture-Agnostic Receiver and Transmitter}
Building upon the proxy-based FL framework, a critical challenge arises from the mismatched feature spaces between the local model $M_i$ and the proxy model $M_{\mathrm{p},i}$. Conventional knowledge transfer techniques~\cite{hinton2015distilling,beyer2022knowledge}, such as conventional parameter averaging, struggle to compute the core interactive logits~$\boldsymbol{z}_{ij}$. To overcome this, we adopt attention-based \emph{knowledge transmitter} and \emph{knowledge receiver} modules, which enable architecture-agnostic knowledge transfer by attention-based feature alignment. The detailed designs are presented as follows.

Before detailing the attention operators, we clarify the model decomposition. The local model $M_i$ comprises a feature extraction body $L_i^b(\cdot)$ and a prediction head $H_i^l(\cdot)$. Similarly, the proxy model $M_{\mathrm{p},i}$ is decomposed into a body $M_{\mathrm{p},i}^b(\cdot)$ and a head $H_i^p(\cdot)$. Knowledge exchange between local models and proxy models is performed exclusively at the body level. 

\subsubsection{Knowledge Receiver}
The knowledge receiver integrates proxy features into the local model during the knowledge injection stage. Let $I_{\mathrm{loc}} \in \mathbb{R}^{N \times D_{\mathrm{loc}}}$ and $I_{\mathrm{p}} \in \mathbb{R}^{N \times D_{\mathrm{p}}}$ denote the intermediate feature representations extracted by the local model body and the proxy model body, respectively. To align heterogeneous feature dimensions, $I_{\mathrm{loc}}$ is first projected into the proxy feature space via a linear mapping function:
\begin{equation}
I'_{\mathrm{loc}} = \phi(I_{\mathrm{loc}}; W_d) = I_{\mathrm{loc}} W_d,
\label{rec3}
\end{equation}
where $W_d \in \mathbb{R}^{D_{\mathrm{loc}} \times D_{\mathrm{p}}}$ maps local features to the proxy feature dimension $D_{\mathrm{p}}$.

Subsequently, the receiver formulates the local representation as queries, and the proxy features as keys and values. By explicitly substituting the projected feature $I'_{\mathrm{loc}}$, the attention components are generated as:
\begin{equation}
Q_R = I'_{\mathrm{loc}} W_q, \quad K_R = I_{\mathrm{p}} W_k, \quad V_R = I_{\mathrm{p}} W_v,
\label{rec1}
\end{equation}
where $W_q, W_k, W_v \in \mathbb{R}^{D_{\mathrm{p}} \times D_{\mathrm{p}}}$ are the trainable projection matrices. Hence, the cross-attention matrix $M_R$, which captures the semantic correlations between local queries and proxy keys can be given as
\begin{equation}
M_R(I_{\mathrm{loc}}, I_{\mathrm{p}}) = \mathrm{Softmax}\!\left( \frac{(I_{\mathrm{loc}} W_d W_q)(I_{\mathrm{p}} W_k)^\top}{\sqrt{D_{\mathrm{p}}}} \right).
\label{rec2}
\end{equation}

Finally, to strictly preserve the intrinsic feature topology, we employ a residual connection. By sequentially substituting~\eqref{rec3},~\eqref{rec1} and~\eqref{rec2} into the residual framework, 
the complete injected local feature $I_{\mathrm{loc},R}$ and the local logit $\boldsymbol{z}^l$ can be explicitly formulated as 
\begin{equation}
\begin{aligned}
I_{\mathrm{loc},R} &= I_{\mathrm{loc}} + \mathrm{Softmax}\!\left( \frac{Q_R K_R^\top}{\sqrt{D_{\mathrm{p}}}} \right) V_R W_{\mathrm{out}} \\
&= I_{\mathrm{loc}} + \mathrm{Softmax}\!\left( \frac{(I_{\mathrm{loc}} W_d W_q)(I_{\mathrm{p}} W_k)^\top}{\sqrt{D_{\mathrm{p}}}} \right)(I_{\mathrm{p}} W_v) W_{\mathrm{out}},\\
\end{aligned}
\end{equation}
where $W_{\mathrm{out}} \in \mathbb{R}^{D_{\mathrm{p}} \times D_{\mathrm{loc}}}$ restores the feature dimension back to the local space, and the local logit $\boldsymbol{z}^l$ can be explicitly formulated as
\begin{equation}
\boldsymbol{z}^l = H_i^l(I_{\mathrm{loc},R})
\end{equation}

\subsubsection{Knowledge Transmitter}
Symmetrically, during the distillation stage, the transmitter extracts localized knowledge into the proxy. Following the dimensional alignment principle via a projection matrix $W'_d$, the proxy feature acts as the foundational query, while aligned local features serve as keys and values. The components are explicitly formulated as:
\begin{equation}
Q_T = I_{\mathrm{p}} W'_q, \quad K_T = (I_{\mathrm{loc}} W'_d) W'_k, \quad V_T = (I_{\mathrm{loc}} W'_d) W'_v.
\end{equation}
The extraction attention matrix $M_T$, evaluating the importance of different local features with respect to the proxy, is derived by substitution:
\begin{equation}
M_T(I_{\mathrm{p}}, I_{\mathrm{loc}}) = \mathrm{Softmax}\!\left( \frac{(I_{\mathrm{p}} W'_q)(I_{\mathrm{loc}} W'_d W'_k)^\top}{\sqrt{D_{\mathrm{p}}}} \right).
\end{equation}
By integrating the attention mechanism with a residual connection and completely substituting the intermediate variables, the distilled proxy feature $I_{\mathrm{p},T}$ is given as
\begin{equation}
\begin{split}
I_{\mathrm{p},T} &= I_{\mathrm{p}} + \mathrm{Softmax}\!\left( \frac{Q_T K_T^\top}{\sqrt{D_{\mathrm{p}}}} \right) V_T \\
&= I_{\mathrm{p}} + \mathrm{Softmax}\!\left( \frac{(I_{\mathrm{p}} W'_q)(I_{\mathrm{loc}} W'_d W'_k)^\top}{\sqrt{D_{\mathrm{p}}}} \right) (I_{\mathrm{loc}} W'_d W'_v),
\end{split}
\end{equation}
and the corresponding proxy logit $\boldsymbol{z}^p$ is explicitly given as
\begin{equation}
\boldsymbol{z}^p = H_i^p(I_{\mathrm{p},T}).
\end{equation}
Through this step-by-step formulation, the proxy model effectively absorbs heterogeneous local knowledge with negligible overhead, as the knowledge exchange operates exclusively on intermediate representations.

\begin{figure}[t]
\centerline{\includegraphics[width=0.4\textwidth]{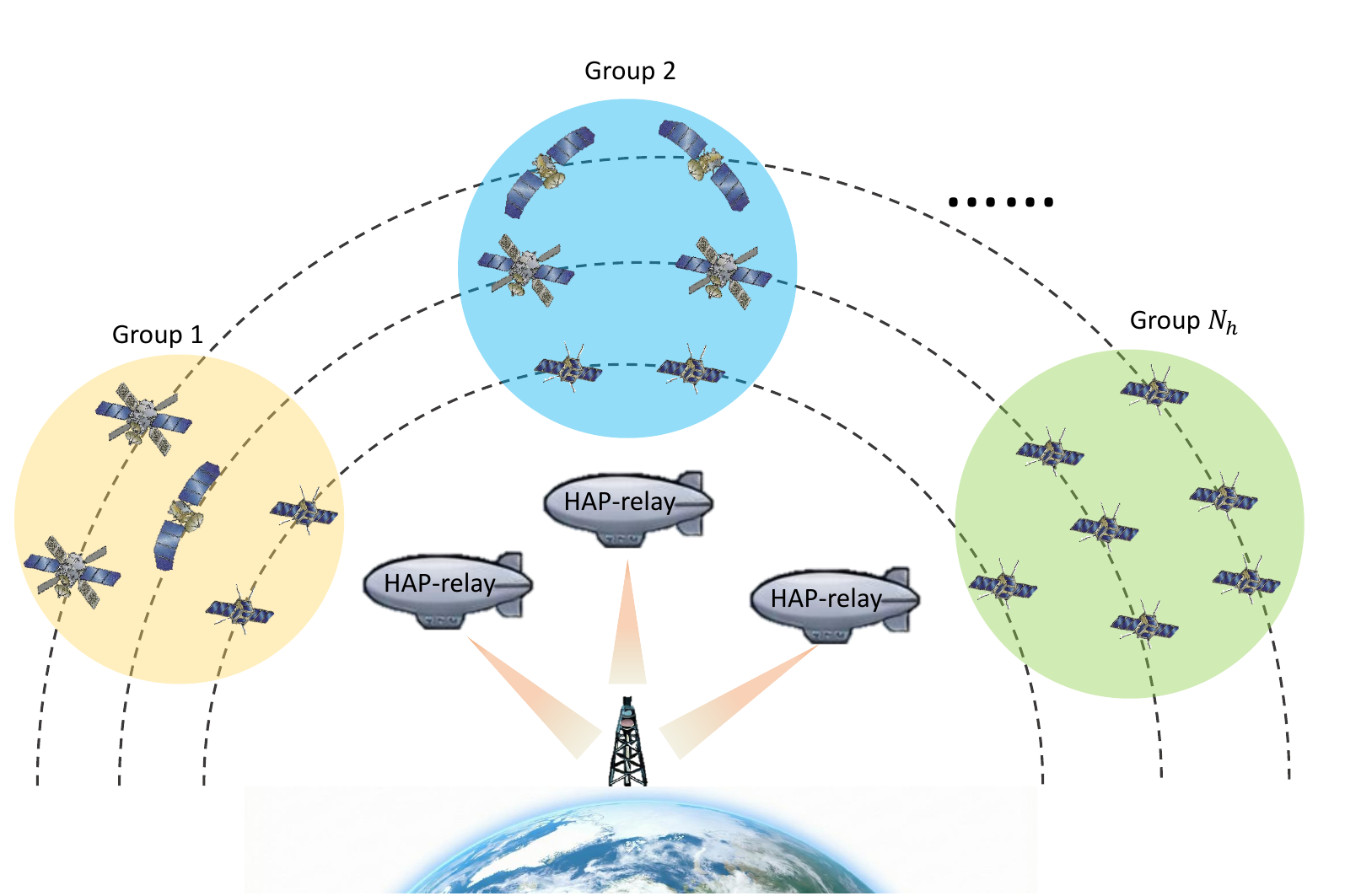}}
\captionsetup{font={footnotesize}}
\caption{Topology-Induced Heterogeneous Satellite Grouping.}
\label{fig:group}
\end{figure}

\section{Topology-Aware Two-Stage Aggregation Mechanism}
\label{aggregation}
While proxy models address architectural heterogeneity, traditional one-stage aggregation still faces an intractable trade-off: synchronous approaches suffer from severe straggler effects due to intermittent contacts, whereas fully asynchronous methods inevitably lead to model staleness that degrades global convergence. To tackle this challenge with physical orbital dynamics, we propose a topology-aware two-stage aggregation mechanism operating over continuous absolute time $t$. Satellites are first dynamically clustered based on their serving HAPs. Then Stage I employs asynchronous aggregation at HAPs to absorb local topological delays, while Stage II enforces synchronous global aggregation at the GS to bound model staleness. The detailed design of the aggregation mechanism is presented as follows.

\subsection{Topology-Induced Heterogeneous Satellite Grouping}
As shown in Fig.~\ref{fig:group}, the multi-layer LEO-HAP-GS network inherently forces satellites to relay updates via visible HAPs, naturally inducing a grouped topology. We mathematically formalize this association over absolute time via a dynamic, hardware-aware spatio-temporal matching process.

Let $\mathcal{A}_m(t)$ denote the active set of satellites associated with HAP $m$ at continuous time $t$.
To optimally balance the transient connectivity and the actual relay processing bottleneck, each satellite $i$ dynamically evaluates a joint utility functional $\mathcal{U}_{i,m}(t)$ given as
\begin{equation}
\mathcal{U}_{i,m}(t) = \int_{\mathcal{T}_{\mathrm{vis}}^{i,m}} \mathcal{C}_{i,m}(\tau) \, d\tau - \mu \sum_{j \in \mathcal{A}_m(t)} \frac{\Phi_{\mathrm{p},j}}{\mathcal{B}_j},
\label{scability_1}
\end{equation}
where the integral computes the exact definable sub-THz channel capacity over the continuous visibility window $\mathcal{T}_{\mathrm{vis}}^{i,m}$. Crucially, the penalty term $\sum_{j \in \mathcal{A}_m(t)} \frac{\Phi_{\mathrm{p},j}}{\mathcal{B}_j}$ explicitly quantifies the true effective workload of the HAP, driven by the proxy model size $\Phi_{\mathrm{p},j}$ and the heterogeneous capability budget $\mathcal{B}_j$ of each associated satellite. $\mu > 0$ is a balancing coefficient.
Satellite $i$ then dynamically associates with the optimal HAP:
\begin{equation}
\mathrm{HAP}_i^\ast(t) = \arg\max_m \mathcal{U}_{i,m}(t).
\end{equation}
Following this association, the $N_s$ satellites are partitioned into $N_h$ time-varying groups, defined as 
\begin{equation}
\mathcal{A}_h = \{ i \mid \text{HAP}_i^\ast = h \}, \quad h = 1, \dots, N_h.  
\end{equation}
Importantly, driven by orbital mechanics and the phased deployment of hardware, these groups encapsulate severe multidimensional heterogeneity. By defining a unified satellite state vector $\boldsymbol{\Theta}_i \triangleq [\boldsymbol{\rho}_i^\top, \mathcal{B}_i, \Phi_{i}]^\top$, this heterogeneity is rigorously quantified as 
\begin{itemize}
\item \textbf{Inter-group size imbalance:} Different HAPs may serve varying numbers of satellites, i.e.,
\begin{equation}
|\mathcal{A}_h| \neq |\mathcal{A}_k|, \quad h \neq k.  
\end{equation}

\item \textbf{Intra-group heterogeneity:} Satellites within the exact same group $h$ exhibit strictly divergent physical capabilities and architectural scales:
\begin{equation}
\quad \boldsymbol{\Theta}_i \neq \boldsymbol{\Theta}_j, \quad \forall i,j \in \mathcal{A}_h(t), \, i \neq j.
\end{equation}
\item \textbf{Inter-group heterogeneity:} The overall physical and computational profile of each group diverges significantly. Defining the macro-level centroid vector $\bar{\boldsymbol{\Theta}}_h(t)$ as the average state of group $h$:
\begin{equation}
\quad \bar{\boldsymbol{\Theta}}_h(t) = \frac{1}{|\mathcal{A}_h(t)|} \sum_{i \in \mathcal{A}_h(t)} \boldsymbol{\Theta}_i,
\end{equation}
\begin{equation}
\quad \bar{\boldsymbol{\Theta}}_h(t) \neq \bar{\boldsymbol{\Theta}}_k(t), \quad \forall h \neq k.
\end{equation}
\end{itemize}

These definitions provide a clear quantitative description of group heterogeneity, which will be directly used in the following design of Stage I and Stage II aggregation procedures.

\subsection{Stage I: Asynchronous Intra-Group Aggregation}
Given the topology-induced heterogeneous groups, Stage I performs asynchronous proxy aggregation at the serving HAPs to instantly absorb satellite updates. To maximize the utilization of transient physical contacts and strictly avoid straggler bottlenecks, this stage abandons rigid round-based synchronization in favor of an event-driven temporal mechanism mapped to the absolute time $t$. 

Let $\boldsymbol{\omega}_h^{t}$ denote the current proxy model maintained by HAP $h$ at time $t$, and let $\boldsymbol{\omega}_{i,h}^{t}$ denote the proxy model uploaded by satellite $i$ from group $h$. 
We formally define an arrival event $\mathcal{E}_{i,h}(t_e)$ as the moment satellite $i \in \mathcal{A}_h(t_e)$ successfully uploads  $\boldsymbol{\omega}_{i,h}^{t}$ to HAP $h$ within its visibility window $t_e \in \mathcal{T}_{\mathrm{vis}}^{i,h}$. Upon the trigger of $\mathcal{E}_{i,h}(t_e)$, the HAP instantaneously executes an incremental aggregation update expressed as
\begin{equation}
\boldsymbol{\omega}_h(t_e^+) = \boldsymbol{\omega}_h(t_e^-) + \eta_{i,h}(t_e) \big( \boldsymbol{\omega}_{i}(t_e) - \boldsymbol{\omega}_h(t_e^-) \big),
\label{stage_I_update}
\end{equation}
where $\eta_i$ is an aggregation weight that controls the contribution of satellite $i$. To explicitly address both model staleness and the intra-group heterogeneity described above, we design the aggregation weight $\eta_i$ as
\begin{equation}
\eta_{i,h} = \frac{\mathcal{B}_i}{\sum_{j \in \mathcal{A}_h(t_e)} \mathcal{B}_j} \cdot \exp \left( - \int_{t_{i,\mathrm{gen}}}^{t_e} \nu_i(\tau) \, d\tau \right),
\label{aggregation_weight}
\end{equation}
In this formulation, the first term $\frac{\mathcal{B}_i}{\sum_{j \in \mathcal{A}_h(t_e)} \mathcal{B}_j}$ reflects the relative hardware-aware contribution of satellite $i$ within the group $h$ considering its intra-group heterogeneity $\mathcal{B}_i$, while $\exp \Big( -\gamma \big(t_e - t_{i,\mathrm{gen}}\big) \Big)$ captures temporal decay during the delay period, where $t_{i,\mathrm{gen}}$ is the exact physical timestamp when satellite $i$ initiated its local training epoch, and $\nu_i(\tau) > 0$ denotes the instantaneous orbital volatility function capturing the time-varying network instability.

By explicitly substituting the temporal decay and the hardware-aware contribution metrics into the event-driven update framework, the complete, end-to-end asynchronous aggregation executed at HAP $h$ is mathematically expanded as
\begin{equation}
\begin{aligned}
\boldsymbol{\omega}_h(t_e^+) &= \left[ \frac{\mathcal{B}_i}{\sum_{j \in \mathcal{A}_h(t_e)} \mathcal{B}_j} \cdot \exp \left( - \int_{t_{i,\mathrm{gen}}}^{t_e} \nu_i(\tau) \, d\tau \right) \right] \\
&\times\Big( \boldsymbol{\omega}_{i}(t_e) - \boldsymbol{\omega}_h(t_e^-) \Big) + \boldsymbol{\omega}_h(t_e^-).
\end{aligned}
\end{equation}
This design guarantees a fair aggregation that resists both temporal staleness and hardware dominance, successfully bounding local drift.

\subsection{Stage II: Inter-Group Synchronous Aggregation}
Following the asynchronous intra-group aggregation in Stage I, the GS executes synchronous aggregation across all HAPs to ensure timely and accurate global knowledge sharing. Since the HAPs maintain persistent links with the GS, this stage operates strictly at predefined absolute physical time intervals $T_{\mathrm{sync}}$. At each discrete synchronization epoch $t_k = k \cdot T_{\mathrm{sync}}$ ($k=1,2,\dots$), the GS receives all the proxy models $\{\boldsymbol{\omega}_h(t_k^-)\}_{h=1}^{N_h}$ to reconstruct the optimal global state $\boldsymbol{\omega}_{\mathrm{G}}(t_k)$.
To address inter-group heterogeneity, we formulate the global aggregation as an Information-Weighted Proximal Objective. 
Specifically, we first define the Accumulated Information Yield $\Gamma_h(t_k)$ for each group $h$ over the preceding interval $[t_{k-1}, t_k]$ given as
\begin{equation}
\begin{aligned}
\Gamma_h(t_k) &= \ln\big(\mathrm{Tr}(\mathbf{I}_h) + 1\big) \\
& \times \int_{t_{k-1}}^{t_k} \sum_{i \in \mathcal{A}_h(\tau)} \Big( |\mathcal{D}_i| \cdot \mathcal{B}_i(\tau) \cdot e^{-\kappa(t_k - \tau)} \Big) \, d\tau,
\end{aligned}
\label{scability_2}
\end{equation}
where $\mathrm{Tr}(\mathbf{I}_h)$ denotes the trace of the empirical Fisher Information Matrix, which explicitly rewards high-capacity architectures, while the $[t_{k-1}, t_k]$ integral continuously accumulates the local data scale $|\mathcal{D}_i|$ coupled with the transient hardware budget $\mathcal{B}_i(\tau)$ under the absolute temporal decay $e^{-\kappa(t_k - \tau)}$.
The GS then solves the following joint optimization problem to find the global equilibrium
\begin{equation}
\min_{\boldsymbol{\omega}} \mathcal{J}(\boldsymbol{\omega}) = \sum_{h=1}^{N_h} \Gamma_h(t_k) \big\| \boldsymbol{\omega} - \boldsymbol{\omega}_h(t_k^-) \big\|_2^2 + \frac{\mu}{2} \big\| \boldsymbol{\omega} - \boldsymbol{\omega}_{\mathrm{G}}(t_{k-1}) \big\|_2^2,
\end{equation}
where the proximal term ($\mu > 0$) ensures global stability by restricting excessive deviations from the prior global state. By setting the first-order optimality condition $\nabla_{\boldsymbol{\omega}} \mathcal{J} = \mathbf{0}$, the synchronous aggregation rule is rigorously derived as the following closed-form solution
\begin{equation}
\boldsymbol{\omega}_{\mathrm{G}}(t_k) = \frac{2 \sum_{h=1}^{N_h} \Gamma_h(t_k) \boldsymbol{\omega}_h(t_k^-) + \mu \boldsymbol{\omega}_{\mathrm{G}}(t_{k-1})}{2 \sum_{m=1}^{N_h} \Gamma_m(t_k) + \mu}.
\end{equation}
This formulation transforms empirical averaging into a mathematically grounded optimization. It guarantees that the global model converges towards the most informative group manifolds while effectively mitigating the staleness-induced drift inherent in dynamic satellite topologies.

\subsection{Theoretical Convergence Analysis}
To mathematically guarantee the global convergence of the proposed topology-aware two-stage aggregation framework, we establish the convergence bound of the proposed mechanism. Since the synchronous aggregation in Stage~II is inherently stable, the critical challenge lies in the asynchronous updates of Stage~I.

First, we map the continuous-time events $t_e$ to a discrete update sequence indexed by $k \in \{1, \dots, K\}$. The asynchronous update rule at the HAP from~\eqref{stage_I_update} can be abstracted as 
\begin{equation}
\omega_{k+1} = \omega_k - \eta_k g_{i_k}(\omega_{k-\tau_k}), 
\end{equation}
where $\eta_k$ corresponds to the weight $\eta_{i,h}(t_e)$ in~\eqref{aggregation_weight}, and $\tau_k \le \tau_{max}$ represents the bounded index delay caused by physical topology blind zones. 

For the theoretical analysis, we adopt standard conditions: the global objective function $F(\omega)$ is $L$-smooth, and the local stochastic gradients $g_i(\omega)$ are unbiased with a bounded variance $\sigma^2$. We first characterize the single-step descent behavior bounded by the topology-induced staleness.

\it
Lemma 1 (Descent Inequality with Staleness): 
Assuming the learning rate satisfies $\eta_k \leq \frac{1}{2L}$, the expected objective function value at the $(k+1)$-th event is bounded by:
\begin{equation}
\begin{aligned}
\mathbb{E}[F(\omega_{k+1})] &\leq \mathbb{E}[F(\omega_k)] - \frac{\eta_k}{2} \mathbb{E} \|\nabla F(\omega_k)\|^2 \\
&+ \frac{\eta_k}{2} \mathbb{E} \|\nabla F(\omega_k) - \nabla F(\omega_{k-\tau_k})\|^2 + \frac{L\eta_k^2}{2}\sigma^2.
\end{aligned}
\end{equation}

Proof: 
\rm By applying the $L$-smoothness expansion $\mathbb{E}[F(\omega_{k+1})] \leq \mathbb{E}[F(\omega_k)] - \mathbb{E} \langle \nabla F(\omega_k), \eta_k g_{i_k}(\omega_{k-\tau_k}) \rangle + \frac{L}{2} \mathbb{E} ||\eta_k g_{i_k}(\omega_{k-\tau_k})||^2$, utilizing the algebraic identity $2\langle a, b \rangle = ||a||^2 + ||b||^2 - ||a-b||^2$, and substituting the bounded variance condition $\mathbb{E}||g||^2 \leq \mathbb{E}||\nabla F||^2 + \sigma^2$, the cross-term can be decoupled to yield the inequality. 
$\hfill\blacksquare$

Building upon Lemma~1, we establish the global convergence bound over $K$ asynchronous events.

\it 
Theorem 1 (Convergence Bound):
Under the maximum topological staleness $\tau_{max}$, the expected gradient norm of the objective function after $K$ asynchronous updates satisfies:
\begin{equation}
\begin{aligned}
\frac{1}{K} \sum_{k=1}^{K} \mathbb{E} \left[ \|\nabla F(\omega_k)\|^2 \right] &\leq \frac{2(F(\omega_1) - F^*)}{\tilde{\eta} K} \\
&+ \tilde{\eta} L \sigma^2 + \frac{2 L^2 \sigma^2}{K} \sum_{k=1}^{K} \eta_k^2 \tau_{max}^2,
\end{aligned}
\end{equation}
where $\tilde{\eta} = \frac{1}{K}\sum_{k=1}^K \eta_k$ is the effective average step size, and $F^*$ is the optimal global objective.

Proof: 
\rm The critical perturbation in Lemma 1 is the staleness error $\mathbb{E} ||\nabla F(\omega_k) - \nabla F(\omega_{k-\tau_k})||^2$. By applying the Cauchy-Schwarz inequality and $L$-smoothness, this divergence is strictly bounded by the accumulated delayed updates:
\begin{equation}
\begin{aligned}
\mathbb{E} \|\nabla F(\omega_k) - &\nabla F(\omega_{k-\tau_k})\|^2 \leq L^2 \mathbb{E} \| \sum_{j=k-\tau_k}^{k-1} \eta_j g_{i_j} \|^2 \\
&\leq L^2 \tau_{max}^2 \max_j (\eta_j^2) (\mathbb{E}\|\nabla F\|^2 + \sigma^2).
\end{aligned}
\end{equation}
Substituting this bound back into Lemma 1, summing the inequality over all events $k = 1$ to $K$ (telescoping sum), and rearranging the terms to isolate the gradient norm $\mathbb{E}||\nabla F(\omega_k)||^2$ directly yields the final bound. 
$\hfill\blacksquare$

Theorem 1 explicitly addresses the critical divergence risk of asynchronous FL. The global convergence is strictly dominated by the staleness perturbation penalty term $\mathcal{O}(\eta_k^2 \tau_{max}^2)$. By leveraging the topology-aware weighting in~\eqref{aggregation_weight}, we enforce an exponential decay $\exp(-\int \nu_i(\tau)d\tau)$ on the weight $\eta_k$ of severely outdated models. This ensures that the staleness penalty $\sum_{k=1}^{K} \eta_k^2 \tau_{max}^2$ strictly converges towards zero as delay increases, thereby completely neutralizing the asynchronous divergence risk and ensuring stable system-wide convergence.

\section{Numerical Results}
\label{numerical}
In this section, we conduct numerical simulations to evaluate the proposed heterogeneous FL framework. We first demonstrate the topological and capacity advantages of the physical architecture. Subsequently, we validate the superiority of our proxy-model-based FL scheme and topology-aware aggregation mechanism against state-of-the-art (SOTA) baselines.

\begin{table*}[t]
\caption{Satellite visibility and average contact window duration under varying massive constellation sizes and orbital inclinations.}
\label{tab:topology_comprehensive}
\centering
\renewcommand{\arraystretch}{1.3}
\setlength{\tabcolsep}{5pt}
\begin{tabular}{l l c c c c c c c c}
\hline\hline
\multirow{3}{*}{\textbf{Inclination}} & \multirow{3}{*}{\textbf{Architecture}} & \multicolumn{4}{c}{\textbf{Visible Satellites (\% of Total)}} & \multicolumn{4}{c}{\textbf{Average Contact Window (s)}} \\
\cmidrule(lr){3-6} \cmidrule(lr){7-10}
 & & $N_s = 50$ & $N_s = 100$ & $N_s = 150$ & $N_s = 200$ & $N_s = 50$ & $N_s = 100$ & $N_s = 150$ & $N_s = 200$ \\
\hline
\multirow{2}{*}{$10^\circ$} 
 & Sat-GS       & 25 (50.0\%) & 50 (50.0\%) & 75 (50.0\%) & 100 (50.0\%) & 315.2 & 315.7 & 315.2 & 315.5 \\
 & Sat-HAP-GS   & \textbf{42 (84.0\%)} & \textbf{83 (83.0\%)} & \textbf{125 (83.3\%)} & \textbf{167 (83.5\%)} & \textbf{769.9} & \textbf{771.1} & \textbf{769.7} & \textbf{770.4} \\
\hline
\multirow{2}{*}{$40^\circ$} 
 & Sat-GS       & 25 (50.0\%) & 50 (50.0\%) & 75 (50.0\%) & 100 (50.0\%) & 721.6 & 721.2 & 721.4 & 721.8 \\
 & Sat-HAP-GS   & \textbf{33 (66.0\%)} & \textbf{67 (67.0\%)} & \textbf{100 (66.7\%)} & \textbf{133 (66.5\%)} & \textbf{797.1} & \textbf{798.2} & \textbf{797.4} & \textbf{798.0} \\
\hline
\multirow{2}{*}{$70^\circ$} 
 & Sat-GS       & 17 (34.0\%) & 33 (33.0\%) & 50 (33.3\%) & 67 (33.5\%)  & 464.4 & 465.4 & 464.4 & 465.1 \\
 & Sat-HAP-GS   & \textbf{33 (66.0\%)} & \textbf{67 (67.0\%)} & \textbf{100 (66.7\%)} & \textbf{133 (66.5\%)} & \textbf{686.8} & \textbf{686.2} & \textbf{687.5} & \textbf{686.9} \\
\hline\hline
\end{tabular}
\end{table*}

\subsection{Implementation and Experimental Setup}
\label{subsec:simulation_setup}
We implement our proposed framework using Python 3.9 on a high-performance server equipped with four NVIDIA GeForce RTX 4090 GPUs. To model LEO heterogeneity, we assign the widely adopted VGG models, e.g. VGG-19, VGG-16, VGG-13, and VGG-11, to represent four budgets $\mathcal{B}_i \in \{1.0, 0.75, 0.5, 0.25\}$, while a lightweight ResNet-44 serves as the shared proxy model. During the local training process, the batch size is uniformly set to $128$, and the learning rates for the knowledge distillation stage $\eta_{dis}$, knowledge injection stage $\eta_{inj}$, knowledge receiver $\eta_{R}$, and knowledge transmitter $\eta_{T}$ are empirically set as $\{\eta_{dis}, \eta_{inj}, \eta_{R}, \eta_{T}\} = \{10^{-4}, 10^{-3}, 10^{-3}, 10^{-3}\}$. The loss functions $L_{dis}$ and $L_{inj}$ are the standard Cross Entropy loss.

We evaluate our proposed framework using the widely recognized remote sensing image dataset EuroSAT~\cite{Helber2018Introducing}. It contains 10 distinct categories of land cover images, 21,600 training samples and 5,400 test samples, making it a highly representative of Earth observation tasks. Moreover, to validate the robustness of our framework against data heterogeneity, we conduct experiments under both identically distributed (IID) and non-IID data settings independently. In the IID setting, the training samples are randomly shuffled and evenly partitioned. In the non-IID setting, the training data is sorted by categorical labels and divided into 240 shards before being equally distributed across the satellites.

\subsection{Topological Advantages of Satellite-HAP-GS Architecture}
\label{subsec:topology_advantages}

Before evaluating the learning performance of the proposed federated learning framework, we first validate the topological advantages of the Satellite-HAP-GS integrated network. To quantify these advantages, we consider a representative multi-altitude Walker-like LEO constellation consisting of $P = 6$ orbital planes with an orbital inclination angle $\alpha$. The orbital planes are evenly assigned to three different altitudes $H_s \in \{500, 1000, 1500\}$~km, and an equal number of satellites are uniformly distributed within each plane. Additionally, a GS is located at ($30^\circ$N, $90^\circ$W) on the ground, and two HAPs are deployed at a fixed altitude of 20~km, located at ($15^\circ$N, $120^\circ$W) and ($45^\circ$N, $120^\circ$W), respectively. For reliable link connections, the minimum elevation angle thresholds for both the satellite-HAP links ($\theta_{min}^{SH}$) and HAP-GS links ($\theta_{min}^{HG}$) are strictly set to $5^\circ$. 

Based on the setup, we thoroughly evaluate the framework by varying two key spatial parameters, e.g., the total number of satellites $N_s \in \{50, 100, 150, 200\}$ and the orbital inclination angle $\alpha\in \{53^\circ, 70^\circ, 90^\circ\}$. Table.~\ref{tab:topology_comprehensive} presents the comprehensive comparison of satellite visibility and contact window according to~\eqref{vis_1},~\eqref{vis_2}, and~\eqref{vis_all}.

\subsubsection{Enhancement of Satellite Visibility}
First, we evaluates the instantaneous satellite visibility, which directly dictates the maximum number of participating satellites in FL. As shown on the left of Table.~\ref{tab:topology_comprehensive}, the conventional Satellite-GS links restrict the visibility of the constellation, capturing only $33.3\%\sim 50.0\%$ of satellites. In contrast, the introduction of HAPs significantly broadens the coverage. At a $70^\circ$ inclination, it helps double the visibility ratio to $66.7\%$ across constellation sizes ($N_s = 50, 100, 150, 200$). Moreover, at $\alpha=10^\circ$ and $\alpha=40^\circ$, the visibility ratio reaches $83.3\%$. By improving satellite visibility, the HAP-relay network ensures higher satellite participation and enriches the global proxy model's knowledge diversity.

\begin{figure*}[t]
\centering
\subfigure[]{
\includegraphics[width=0.3\textwidth]{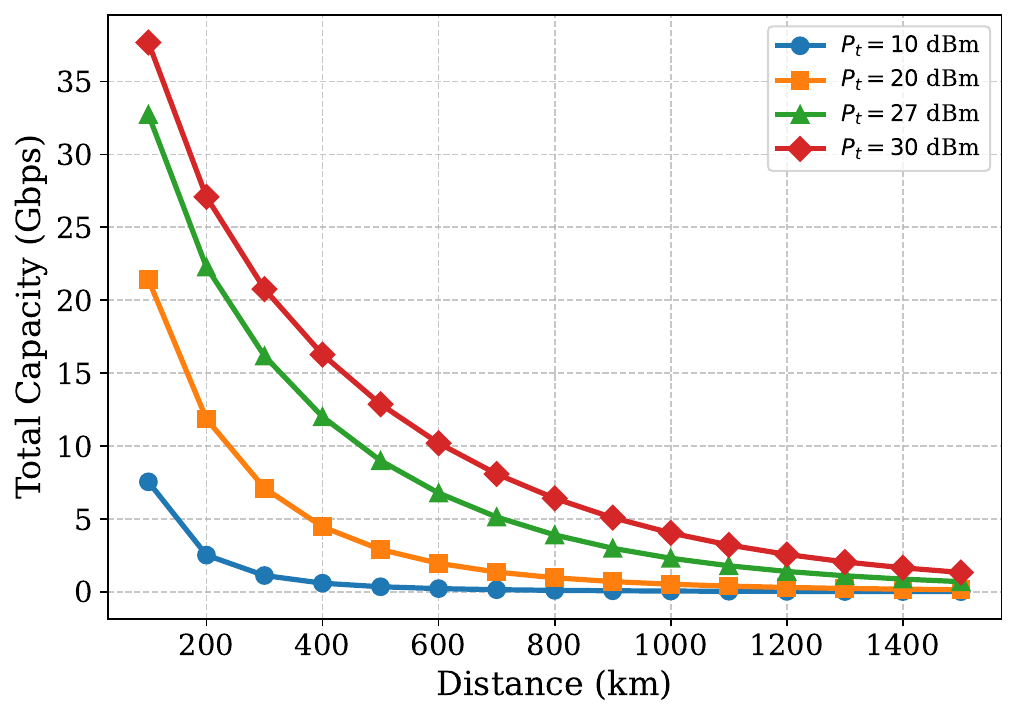}
}
\subfigure[]{
\includegraphics[width=0.3\textwidth]{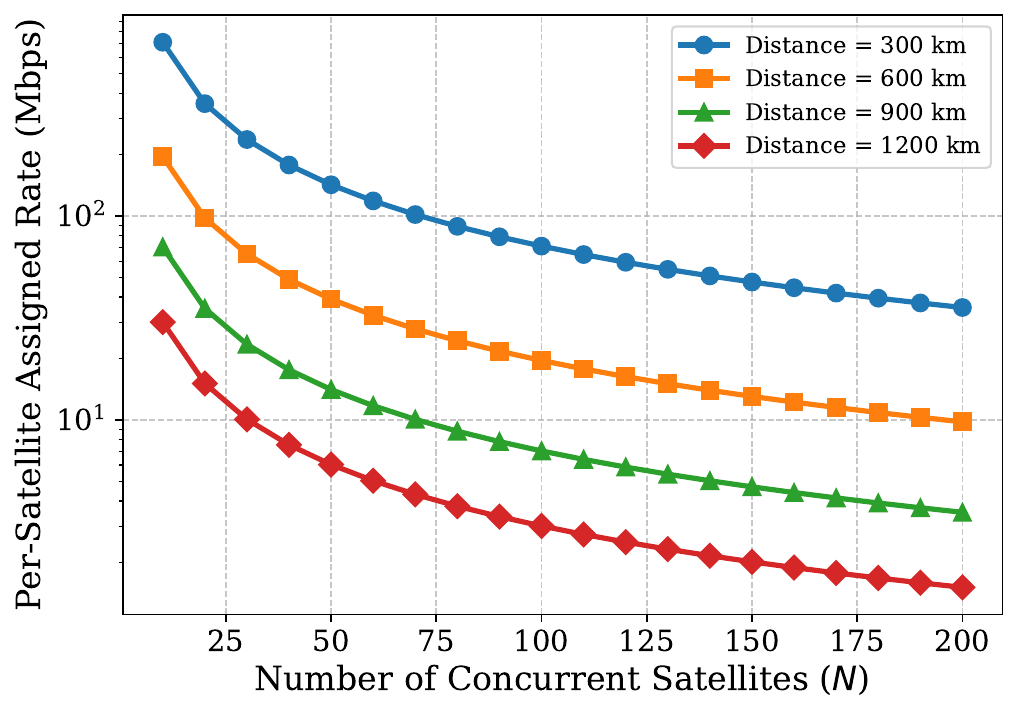}
}
\subfigure[]{
\includegraphics[width=0.3\textwidth]{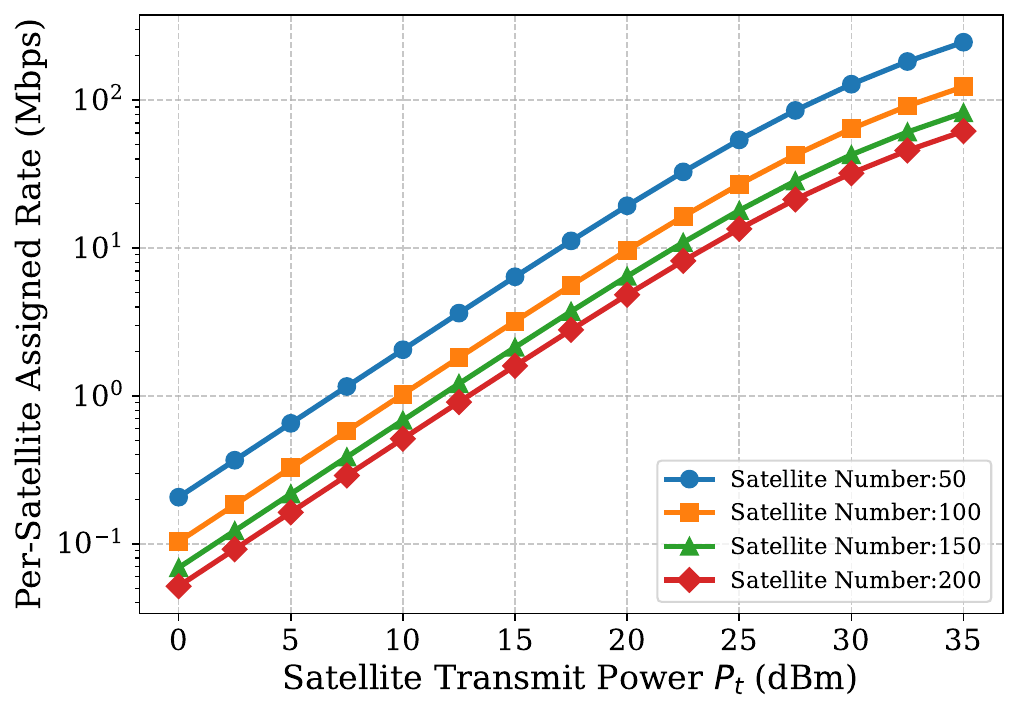}
}
\captionsetup{font={footnotesize}}
\caption{Capacity evaluation of the Sub-THz link. (a) Total capacity versus distance under varying transmit powers; (b) Per-satellite assigned rate versus distance; (c) Per-satellite rate versus transmit power.}
\label{fig_capacity}
\captionsetup{font={footnotesize}}
\end{figure*}

\begin{figure*}[t]
\centering
\subfigure[]{
\includegraphics[width=0.3\textwidth]{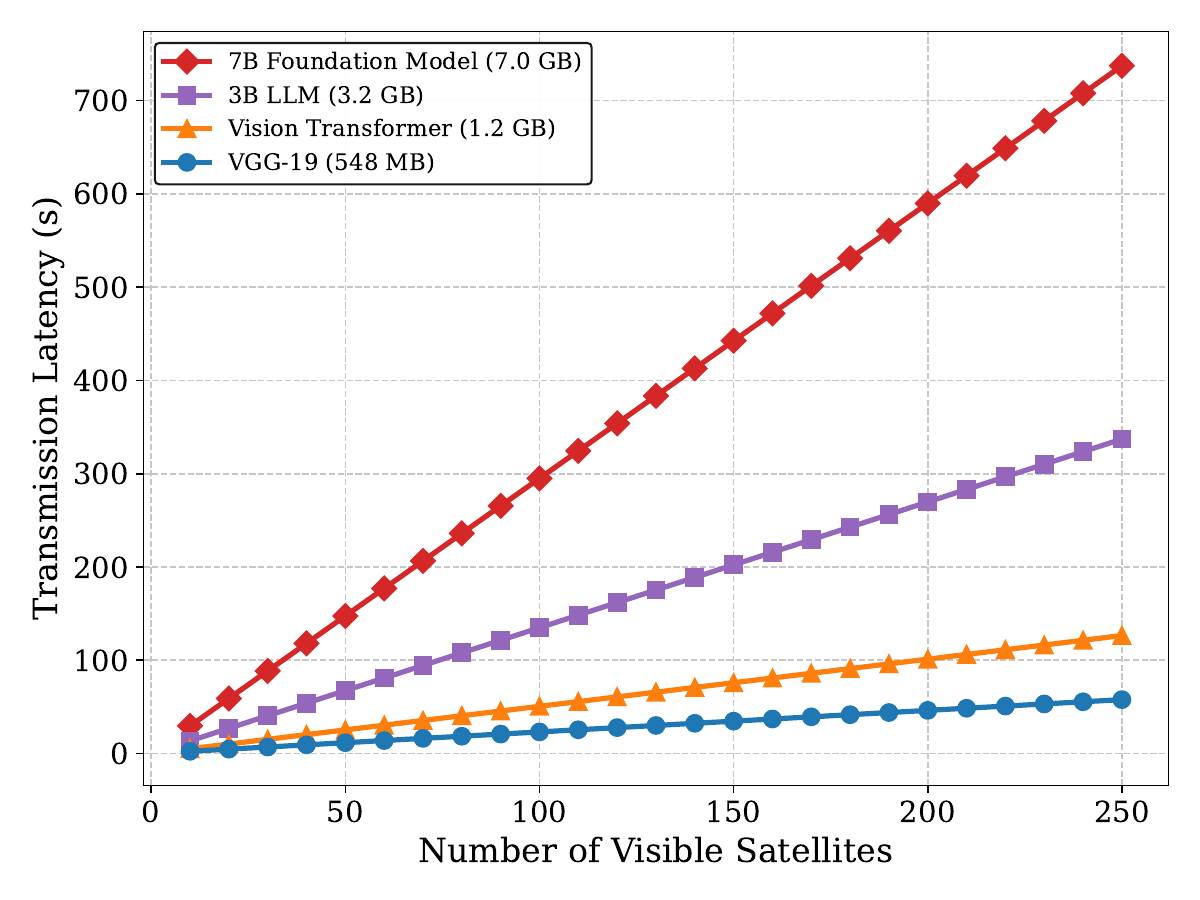}
}
\subfigure[]{
\includegraphics[width=0.3\textwidth]{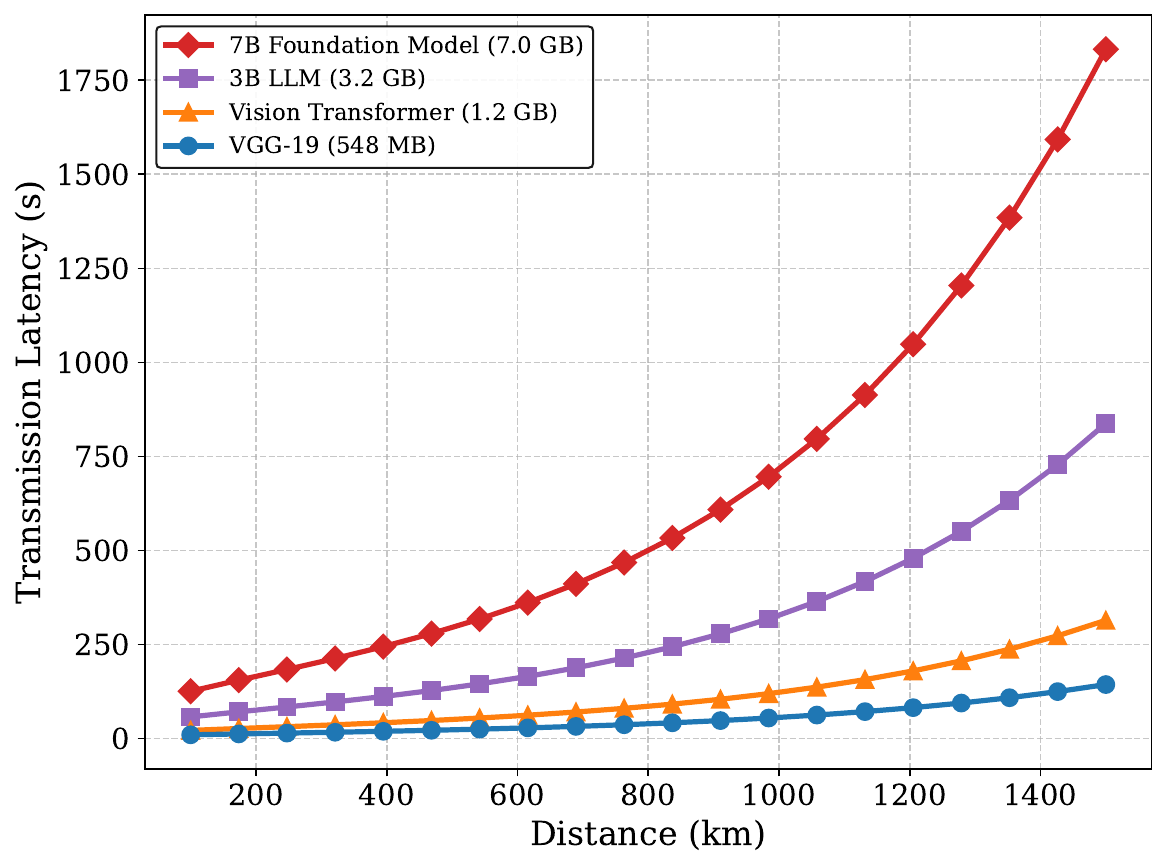}
}
\subfigure[]{
\includegraphics[width=0.3\textwidth]{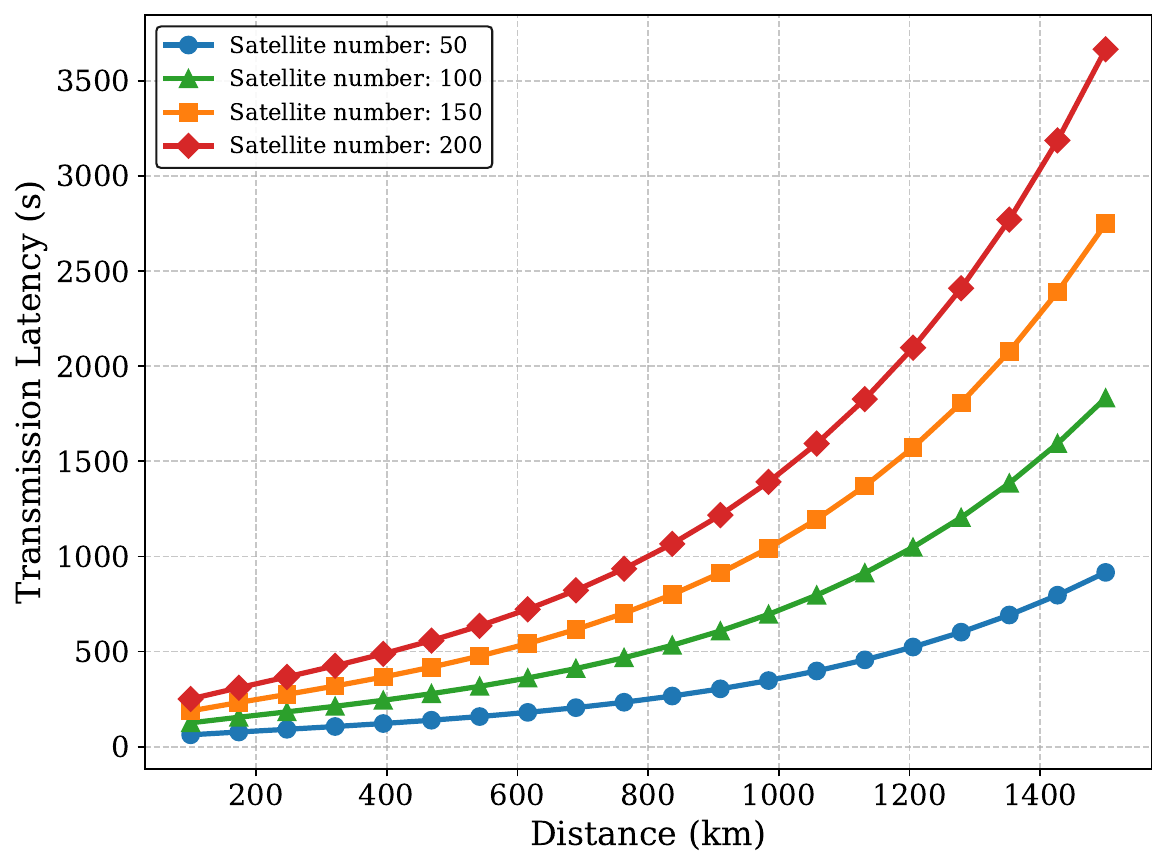}
}
\captionsetup{font={footnotesize}}
\caption{Evaluation of transmission latency over Sub-THz links. (a) Latency versus number of visible satellites at a fixed distance of $800 \text{ km}$; (b) Latency versus distance at different model size; (c) Latency versus distance at different satellites numbers.}
\label{fig_latency}
\captionsetup{font={footnotesize}}
\end{figure*}

\subsubsection{Extension of Communication Contact Windows}
Furthermore, we evaluate the communication contact windows $T_{vis,i,m}$, which directly influence the budget~$\mathcal{B}_i$. The right metric group of Table.~\ref{tab:topology_comprehensive} demonstrates that the average contact window is highly sensitive to $\alpha$. For instance, the Sat-GS links suffer from severely short contact durations of approximately $315\text{ s}$  and $465\text{ s}$ at $10^\circ$ and $70^\circ$ inclinations, respectively. By leveraging the advantages of HAP relays, the proposed architecture yields remarkable temporal extensions to around $700\text{ s}$ at $\alpha=70^\circ$ and nearly $770\text{ s}$ at $\alpha = 10^\circ$, marking an extraordinary improvement of over $50\%\sim140\%$. Even at $\alpha=40^\circ$ with inherently long $T_{vis,i,m}$, the HAPs still provide a steady enhancement from around $720\text{ s}$ to nearly $800\text{ s}$. This extension of communication contact windows effectively reduces stragglers from connection drops, ensuring sufficient bandwidth for satellites to reliably transmit distilled proxy models.

\begin{figure}[t]
\centering
\subfigure[]{
\includegraphics[width=0.4\textwidth]{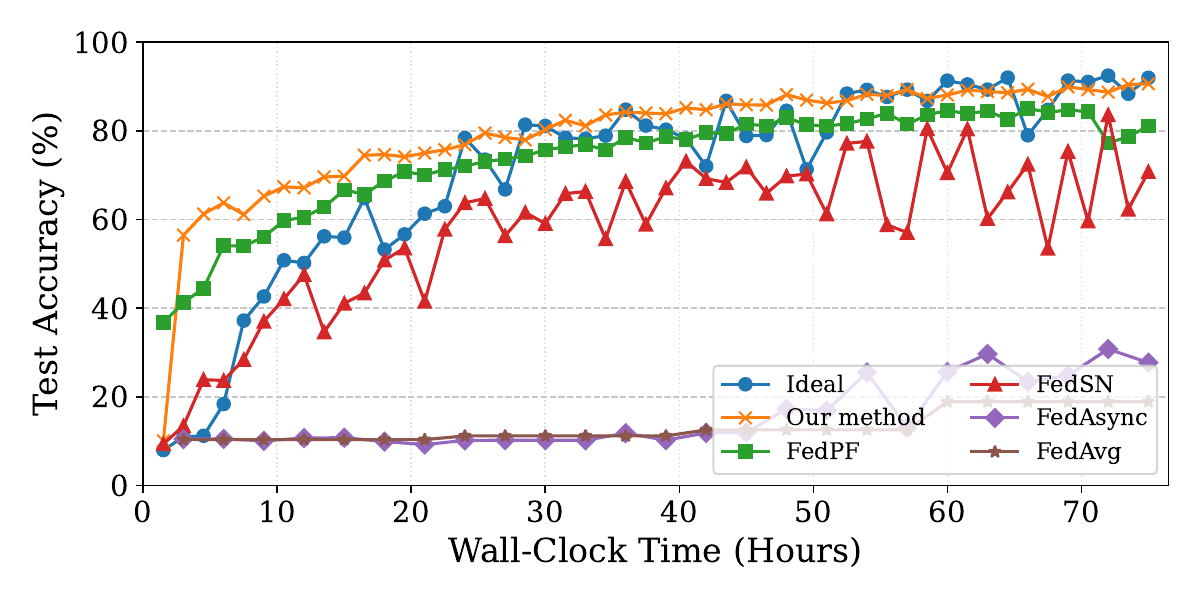}
}
\subfigure[]{
\includegraphics[width=0.4\textwidth]{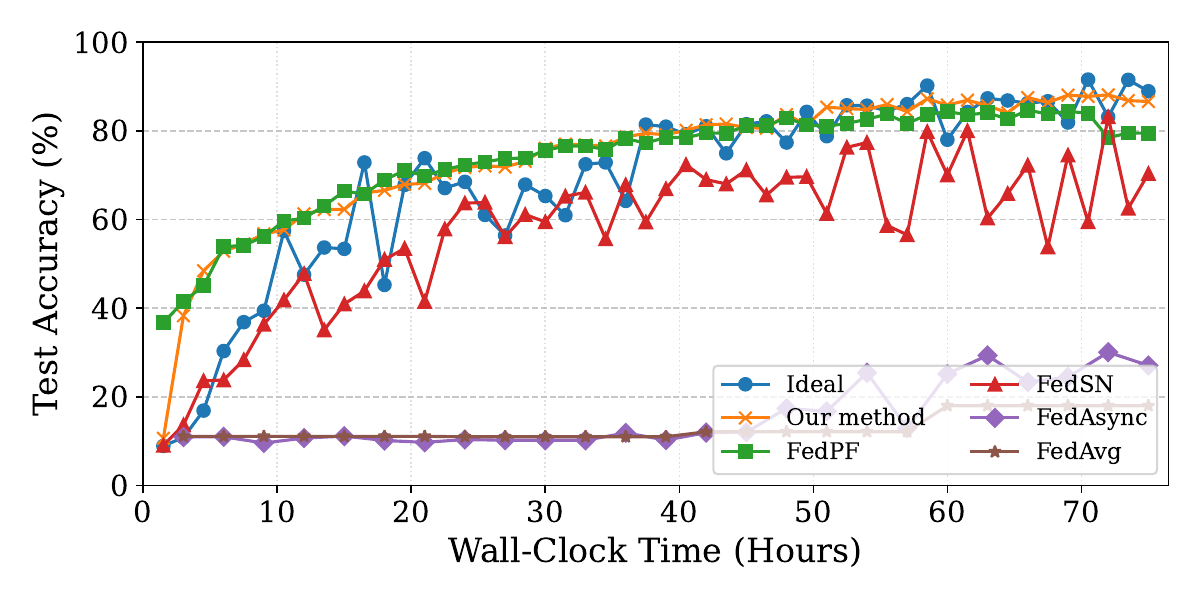}
}
\captionsetup{font={footnotesize}}
\caption{Convergence of test accuracy under (a) the IID data setting; (b) the Non-IID data}
\label{fig:convergence}
\captionsetup{font={footnotesize}}
\end{figure}

\subsection{Capacity Advantages and Physical Feasibility of Sub-THz Links}
\label{subsec:capacity_adv}
Building upon the extended contact windows, we now evaluate the capacity advantages and physical feasibility of Sub-THz Links formulated in Section \ref{system_model}.
To rigorously quantify these advantages, we align the physical parameters with realistic CubeSat capabilities and aerospace channel models~\cite{Masihi2025Terahertz}. Specifically, the satellite-HAP links operate at the FCC-allocated 94.1-100~GHz W-band~\cite{fcc2025spectrum}. The default transmit power is set to $P_t = 20 \text{ dBm}$ ($0.1 \text{ W}$). The antenna diameters for the satellites and HAPs are set to $d_{tx} = 0.2~\text{m}$ and $d_{rx} = 0.5~\text{m}$, respectively. The equivalent stratospheric ambient temperature is set to $T_{atm} = 220\text{ K}$, and the receiver noise figure is $N_f = 10 \text{ dB}$.

\subsubsection{Immense Capacity Advantages}
We evaluate the Sub-THz link capacity from three perspectives. First, fig.~\ref{fig_capacity}(a) illustrates the total capacity reaches $38$ Gbps at $100 \text{ km}$ and maintains several Gbps at an extreme $1500 \text{ km}$. Second, fig.~\ref{fig_capacity}(b) shows that the per-satellite assigned rate remains robust of $100~\text{Mbps}$ at $300 \text{ km}$. Third, fig.~\ref{fig_capacity}(c) shows that a moderate power of 20 dBm (0.1 W) can reach an exceptional data rates of 100 Mbps, strictly adhering to the energy constraints. These results validate that Sub-THz links deliver the ultra-high capacity required for massive satellite access.

\subsubsection{Physical Feasibility}
The physical feasibility of transmission latency within the contact windows is evaluated in Fig.~\ref{fig_latency}. Fig.~\ref{fig_latency}(a) demonstrates that under a favorable distance of $800~\text{km}$), transmitting a massive 7 GB foundation model for 200 satellites requires~$700~\text{s}$, marginally fitting within the extended HAP contact window. However, Fig.~\ref{fig_latency}(b) and Fig.~\ref{fig_latency}(c) reveal that as the distance approaches $1500 \text{ km}$, the transmission latency for the identical 7 GB foundation model and 200 satellites reaches up to $1500\sim3500~\text{s}$. These bottlenecks clearly highlight the necessity of our adapted proxy model-based method, which significantly reduces the communication latency by exchanging a lightweight proxy model rather than the massive full-scale model.

\begin{figure}[t]
\centering
\subfigure[]{
\includegraphics[width=0.4\textwidth]{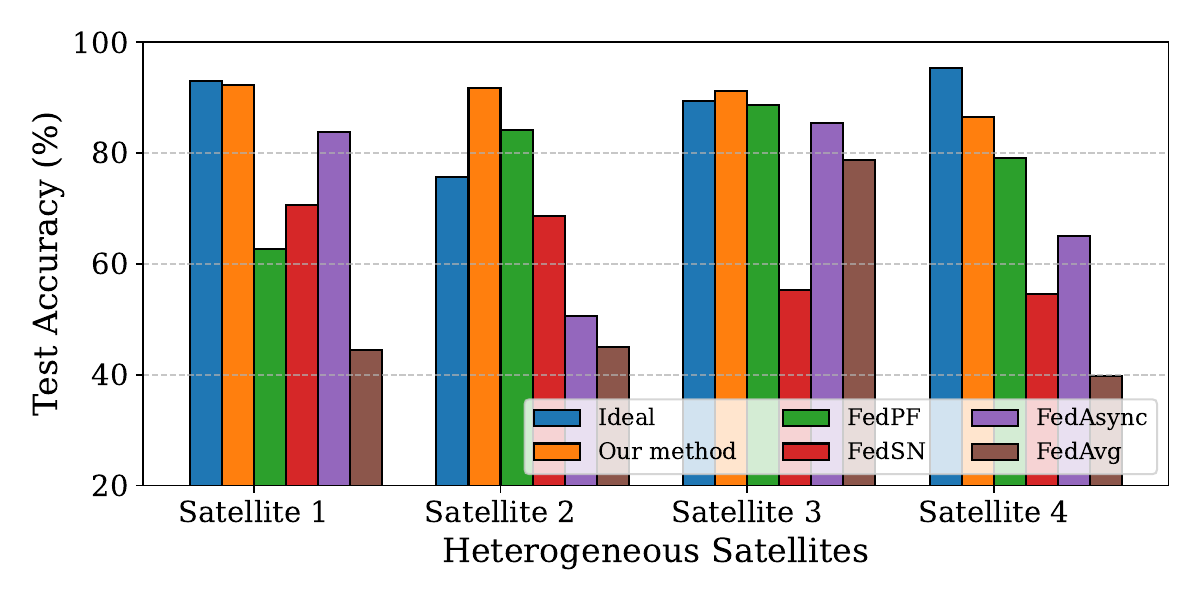}
}
\subfigure[]{
\includegraphics[width=0.4\textwidth]{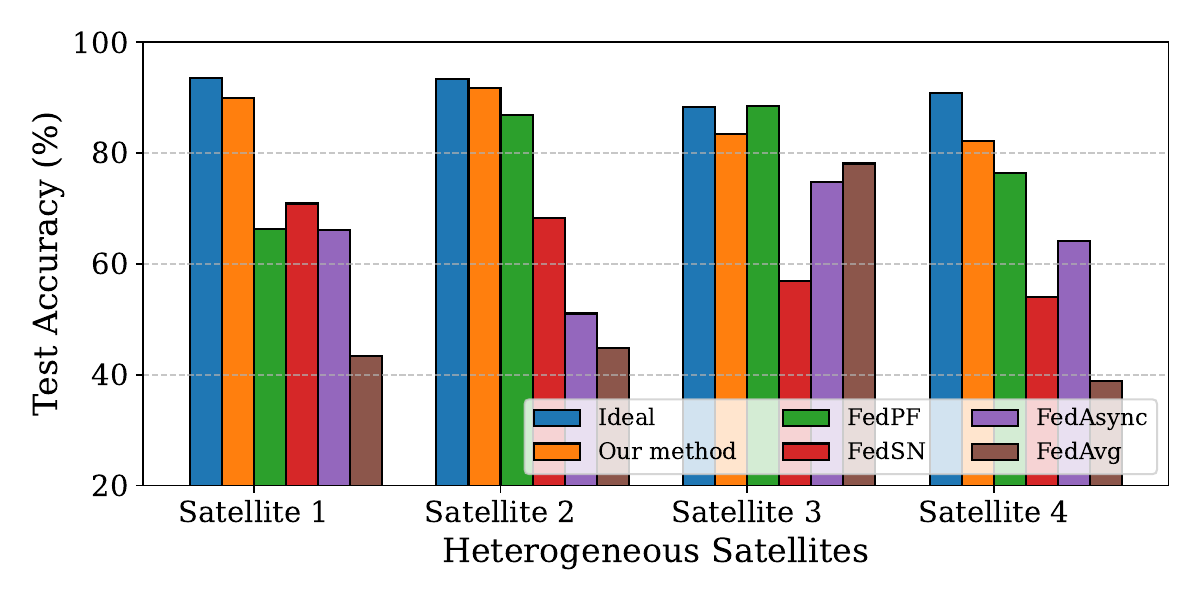}
}
\captionsetup{font={footnotesize}}
\caption{Test accuracy achieved by each individual participating satellite under (a) the IID data setting; (b) the Non-IID data. settings}
\label{fig:accuracy}
\captionsetup{font={footnotesize}}
\end{figure}


\subsection{Accuracy and Convergence Analysis}
\label{subsec:convergence_accuracy}
In this subsection, we evaluate the learning performance of the proposed heterogeneous FL framework. Building upon Sec.~\ref{subsec:topology_advantages} and Sec.~\ref{subsec:capacity_adv}, we adopt a practical simulation scenario: a $70^\circ$ inclined constellation of $N_s = 6$ satellites, where 4 satellites remain visible to the HAP-assisted network for aggregation. Specifically, the 4 participating satellites are assigned distinct resource budgets $b_i \in \{1.0, 0.75, 0.5, 0.25\}$ to capture the heterogeneity. The analytical generalization of this specific setup to massive constellations is explicitly addressed later in this subsection. 
To demonstrate the superiority of the proposed framework, we compare its performance against the following five representative baseline schemes
\begin{itemize}
    \item \textbf{FedAvg:} The standard synchronous federated learning algorithm where all clients share a uniform model architecture~\cite{mcmahan2017communication}.
    \item \textbf{FedAsync:} The standard asynchronous federated optimization framework designed to mitigate the straggler effect~\cite{xie2019asynchronous}.
    \item \textbf{FedSN:} The state-of-the-art heterogeneous federated learning framework specifically designed for LEO satellite networks, which addresses heterogeneity via sub-structure model partitioning~\cite{lin2024fedsn}.
    \item \textbf{FedPF:} The proxy-model-based heterogeneous federated learning baseline. It utilizes the same shared proxy model for architecture-agnostic knowledge exchange as our proposed framework but lacks a specialized aggregation design, relying purely on standard asynchronous aggregation~\cite{xie2024mh}.
    \item \textbf{Ideal:} An ideal, resource- and staleness-unconstrained case of FedAvg. In this theoretical upper bound, it is assumed that all participating satellites possess sufficient computing, storage, and communication resources to perform timely model updates and transmissions for synchronous local model aggregation.
\end{itemize}

\begin{table}[t]
\caption{Average Final Test Accuracy Under IID and Non-IID Settings}
\label{tab:average_accuracy}
\centering
\renewcommand{\arraystretch}{1.3}
\setlength{\tabcolsep}{12pt}
\begin{tabular}{l c c}
\hline\hline
\textbf{FL Scheme} & \textbf{IID} & \textbf{Non-IID} \\
\hline
\textbf{FedAvg}   & 0.5197 & 0.5132 \\
\textbf{FedAsync} & 0.7707 & 0.6652 \\
\textbf{FedSN}    & 0.7077 & 0.7033 \\
\textbf{FedPF}    & 0.8106 & 0.7944 \\

\textbf{Proposed} & \textbf{0.9057} & \textbf{0.8659} \\
\textbf{Ideal}    & 0.9194 & 0.8890 \\
\hline\hline
\end{tabular}
\end{table}

\subsubsection{Accelerated Wall-Clock Convergence Rate} 
Building upon the physical transmission bottlenecks established above, we evaluate global convergence against actual wall-clock time, strictly bounded by the transient HAP contact window $\mathcal{T}_{vis} = 686~\text{s}$ at $\alpha=70^\circ$ shown in Table.~\ref{tab:topology_comprehensive} and~Fig.~\ref{fig_latency}.  
As depicted in Fig.~\ref{fig:convergence}, our method achieves significantly faster and remarkably more stable convergence than all baselines, reaching a 75\% accuracy threshold within only $40\sim45$ hours under the IID setting and non-IID setting, respectively. In contrast, \textbf{FedPF} and \textbf{FedSN} require approximately $60\sim70$ hours, and \textbf{FedAsync} and \textbf{FedAvg} even fail to reach this target after 70 hours. This translates to a convergence speedup of over 1.5$\times$ 2.2$\times$ compared to\textbf{FedPF} and \textbf{FedSN}.

\subsubsection{Superiority in Final Accuracy and Model Heterogeneity} 
As summarized in Table~\ref{tab:average_accuracy}, our proposed method demonstrates high accuracy and exceptional robustness under both IID and Non-IID settings. Specifically, our proposed method surpasses standard baselines \textbf{FedAvg} and \textbf{FedAsync} by up to 38.60\%, and outperform the SOTA heterogeneous method \textbf{FedSN} by $19.80\%\sim16.26\%$ in IID and Non-IID settings, respectively. Furthermore, our method achieves remarkable average accuracies of 90.57\% and 86.59\% in the IID and Non-IID setting, closely approaching the unconstrained \textbf{Ideal} upper bounds (91.94\% and 88.90\%).
These results confirm that our adapted proxy-model mechanism reliably aligns heterogeneous features without strict structural matching, guaranteeing robust global convergence.

\subsubsection{Effectiveness of Two-Stage Topology-Aware Aggregation} 
To validate the Effectiveness of the proposed two-stage aggregation mechanism, we compare our proposed method against the \textbf{FedPF} baseline. While both methods utilize the same shared proxy model, the \textbf{FedPF} baseline relies purely on standard asynchronous aggregation, which serves as a direct ablation study. As shown in Table.~\ref{tab:average_accuracy}, our method outperforms \textbf{FedPF} in test accuracy by approximately 9.51\% under the IID setting and 7.15\% under the Non-IID setting. This improvement confirms that our customized Two-Stage Topology-Aware Aggregation is essential and effective compared to standard asynchronous aggregation. 

\subsubsection{Accuracy Consistency Across Heterogeneous Satellites} 
Beyond average accuracy, our proposed framework explicitly ensures accuracy consistency among the four satellites with varying resource budgets. As illustrated in Fig.~\ref{fig:accuracy}, \textbf{FedAvg}, \textbf{FedAsync} and \textbf{FedSN} suffer from severe performance imbalances across different satellites. For instance, \textbf{FedSN} exhibits extreme inter-satellite accuracy imbalances, ranging from 87.87\% down to 44.96\% under the IID setting, and from 76.19\% down to 51.04\% under the non-IID setting. Conversely, our method tightly bounds the accuracy discrepancy among all four heterogeneous nodes to within a narrow margin of 7.76\% under the Non-IID setting (ranging from 84.43\% to 90.77\%) and merely 7.05\% under the IID setting (ranging from 85.46\% to 92.51\%). 
This consistency validates that our proposed method effectively prevents structural biases, ensuring system-wide robustness without sacrificing the learning capabilities of resource-constrained satellites.

\subsubsection{Scalability and Generalization to Massive Constellations}
While our simulations consider only $N_s = 6$ satellites (4 active) for simplicity, this setup rigorously captures the core heterogeneity by assigning distinct resource budgets. Moreover, scaling to massive constellations (e.g., $N_s \ge 50$) can amplify our advantages.
When under massive concurrent access, degraded per-satellite data rates shown in fig.~\ref{fig_capacity}(a) cause transmission latencies for full-scale models (e.g., 548 MB) 
to surge and persistently violate the contact window $\mathcal{T}_{vis}=686~\text{s}$ in fig.~\ref{fig_latency}(c). Conversely, our proposed framework exchanges a lightweight 2.6 MB proxy model, which requires only a few seconds even under extreme bandwidth contention ($N_s = 200$). This perfectly adheres to physical constraints, guaranteeing scalable and efficient global aggregation. However, it is important to note that the per-HAP computation evaluated in~\eqref{scability_1} and~\eqref{scability_2} also scales with the group size, and analyzing this computational overhead at much larger $N_{s}$ remains an area for future work.

\section{Conclusion}
\label{conclusion}
In this paper, we have proposed a topology-aware two-Stage FL framework for Sub-THz heterogeneous LEO communications to effectively tackle the critical challenges of communication, computational resources, and severe model staleness in large-scale LEO constellations. By deploying HAPs as intermediate relays and utilizing high-capacity Sub-THz links, the proposed architecture fundamentally alleviates communication bottlenecks and substantially extends the satellite visibility windows. To accommodate diverse onboard computational resources, we adapt a proxy-model-based approach to fully utilize heterogeneous resources and enable architecture-agnostic knowledge aggregation. Finally, a topology-aware two-stage aggregation mechanism is developed to effectively balance asynchronous intra-group communication efficiency with synchronous inter-group aggregation accuracy. 
Comprehensive numerical results based on the Walker constellation and the EuroSAT dataset validate the superiority of the proposed framework. By effectively addressing model heterogeneity and communication bottlenecks, our approach yields a remarkable test accuracy of up to 90.57\%, outperforming existing SOTA baselines by 16.26\% to 19.80\%. Moreover, our approach accelerates convergence, achieving up to a 2.2$\times$ speedup. This work provides a scalable and robust foundation for enabling advanced distributed intelligence in the realm of Sub-THz SAGIN.

\bibliographystyle{ieeetr}
\bibliography{ref}

@article{yang20196g,
  title={6{G} wireless communications: Vision and potential techniques},
  author={Yang, Ping and Xiao, Yue and Xiao, Ming and Li, Shaoqian},
  journal={IEEE {N}etwork},
  volume={33},
  number={4},
  pages={70--75},
  year={2019},
  publisher={IEEE}
}

@ARTICLE{11242136,
  author={Han, Chong and Gao, Weijun and Yang, Chuang and Peng, Mugen and Zhang, Wenjun},
  journal={IEEE Wireless Communications}, 
  title={Joint Communication and Radar Sensing for Terahertz Space-Air-Ground Integrated Networks ({SAGIN})}, 
  year={2025},
  volume={},
  number={},
  pages={1-8},
}

@article{gao2025terahertz,
  title={{T}erahertz aerospace communications: enabling technologies and future directions},
  author={Gao, Weijun and Han, Chong and Chen, Yong and He, Yuanzhi and Zhang, Wenjun},
  journal={Science China Information Sciences},
  volume={68},
  number={12},
  pages={220302:1–220302:12},
  year={2025},
}

@article{dang2020should,
  title={What should 6{G} be?},
  author={Dang, Shuping and Amin, Osama and Shihada, Basem and Alouini, Mohamed-Slim},
  journal={Nature Electronics},
  volume={3},
  number={1},
  pages={20--29},
  year={2020},
  publisher={Nature Publishing Group UK London}
}

@article{zhao2019uav,
  title={{UAV}-assisted emergency networks in disasters},
  author={Zhao, Nan and Lu, Weidang and Sheng, Min and Chen, Yunfei and Tang, Jie and Yu, F Richard and Wong, Kai-Kit},
  journal={IEEE Wireless Communications},
  volume={26},
  number={1},
  pages={45--51},
  year={2019},
  publisher={IEEE}
}

@article{zhu2020millimeter,
  title={Millimeter-wave full-duplex {UAV} relay: Joint positioning, beamforming, and power control},
  author={Zhu, Lipeng and Zhang, Jun and Xiao, Zhenyu and Cao, Xianbin and Xia, Xiang-Gen and Schober, Robert},
  journal={IEEE Journal on Selected Areas in Communications},
  volume={38},
  number={9},
  pages={2057--2073},
  year={2020},
  publisher={IEEE}
}

@article{fang2022olive,
  title={Olive branch learning: A topology-aware federated learning framework for space-air-ground integrated network},
  author={Fang, Qingze and Zhai, Zhiwei and Yu, Shuai and Wu, Qiong and Gong, Xiaowen and Chen, Xu},
  journal={IEEE Transactions on Wireless Communications},
  volume={22},
  number={7},
  pages={4534--4551},
  year={2022},
  publisher={IEEE}
}

@article{zhai2023fedleo,
  title={{F}ed{LEO}: An offloading-assisted decentralized federated learning framework for low earth orbit satellite networks},
  author={Zhai, Zhiwei and Wu, Qiong and Yu, Shuai and Li, Rui and Zhang, Fei and Chen, Xu},
  journal={IEEE Transactions on Mobile Computing},
  volume={23},
  number={5},
  pages={5260--5279},
  year={2023},
  publisher={IEEE}
}

@article{ahmmed2022digital,
  title={The digital divide in Canada and the role of {LEO} satellites in bridging the gap},
  author={Ahmmed, Tuheen and Alidadi, Afsoon and Zhang, Zichao and Chaudhry, Aizaz U and Yanikomeroglu, Halim},
  journal={IEEE Communications Magazine},
  volume={60},
  number={6},
  pages={24--30},
  year={2022},
  publisher={IEEE}
}

@article{chen2022robust,
  title={Robust task scheduling for delay-aware IoT applications in civil aircraft-augmented SAGIN},
  author={Chen, Qian and Meng, Weixiao and Han, Shuai and Li, Cheng and Chen, Hsiao-Hwa},
  journal={IEEE Transactions on Communications},
  volume={70},
  number={8},
  pages={5368--5385},
  year={2022},
  publisher={IEEE}
}

@inproceedings{wu2024accelerating,
  title={Accelerating handover in mobile satellite network},
  author={Wu, Jiasheng and Su, Shaojie and Wang, Xiong and Zhang, Jingjing and Gao, Yue},
  booktitle={Proc. of IEEE Conference on Computer Communications (INFOCOM)},
  pages={531--540},
  year={2024},
  organization={IEEE}
}

@inproceedings{yuan2023graph,
  title={Graph learning for multi-satellite based spectrum sensing},
  author={Yuan, Haoxuan and Chen, Zhe and Lin, Zheng and Peng, Jinbo and Fang, Zihan and Zhong, Yuhang and Song, Zihang and Wang, Xiong and Gao, Yue},
  booktitle={2023 IEEE 23rd International Conference on Communication Technology (ICCT)},
  pages={1112--1116},
  year={2023},
  organization={IEEE}
}

@article{liu2021uav,
  title={An {UAV}-enabled intelligent connected transportation system with 6{G} communications for Internet of Vehicles},
  author={Liu, Run and Liu, Anfeng and Qu, Zhenzhe and Xiong, Neal N},
  journal={IEEE Transactions on Intelligent Transportation Systems},
  volume={24},
  number={2},
  pages={2045--2059},
  year={2021},
  publisher={IEEE}
}

@article{liu2021novel,
  title={A novel non-stationary 6{G} {UAV} channel model for maritime communications},
  author={Liu, Yu and Wang, Cheng-Xiang and Chang, Hengtai and He, Yubei and Bian, Ji},
  journal={IEEE Journal on Selected Areas in Communications},
  volume={39},
  number={10},
  pages={2992--3005},
  year={2021},
  publisher={IEEE}
}

@article{huang2024energy,
  title={Energy efficiency maximization in {UAV}-assisted intelligent autonomous transport system for 6{G} networks with energy harvesting},
  author={Huang, Jie and Yu, Tao and Zhu, Xiaogang and Yang, Fan and Lai, Xianzhi and Alfarraj, Osama and Yu, Keping},
  journal={IEEE Transactions on Intelligent Transportation Systems},
  year={2025},
  volume={26},
  number={10},
  pages={17212-17222},
  publisher={IEEE}
}

@article{letaief2021edge,
  title={Edge artificial intelligence for 6{G}: Vision, enabling technologies, and applications},
  author={Letaief, Khaled B and Shi, Yuanming and Lu, Jianmin and Lu, Jianhua},
  journal={IEEE journal on selected areas in communications},
  volume={40},
  number={1},
  pages={5--36},
  year={2021},
  publisher={IEEE}
}

@article{chen2021rf,
  title={{RF}-based human activity recognition using signal adapted convolutional neural network},
  author={Chen, Zhe and Cai, Chao and Zheng, Tianyue and Luo, Jun and Xiong, Jie and Wang, Xin},
  journal={IEEE Transactions on Mobile Computing},
  volume={22},
  number={1},
  pages={487--499},
  year={2021},
  publisher={IEEE}
}

@article{so2022fedspace,
  title={Fedspace: An efficient federated learning framework at satellites and ground stations},
  author={So, Jinhyun and Hsieh, Kevin and Arzani, Behnaz and Noghabi, Shadi and Avestimehr, Salman and Chandra, Ranveer},
  journal={arXiv preprint arXiv:2202.01267},
  year={2022}
}

@inproceedings{vasisht2021l2d2,
  title={{L}2{D}2: Low latency distributed downlink for {LEO} satellites},
  author={Vasisht, Deepak and Shenoy, Jayanth and Chandra, Ranveer},
  booktitle={Proc of the ACM Special Interest Group on Data Communication (SIGCOMM)},
  pages={151--164},
  year={2021}
}

@article{chen2020joint,
  title={A joint learning and communications framework for federated learning over wireless networks},
  author={Chen, Mingzhe and Yang, Zhaohui and Saad, Walid and Yin, Changchuan and Poor, H Vincent and Cui, Shuguang},
  journal={IEEE transactions on wireless communications},
  volume={20},
  number={1},
  pages={269--283},
  year={2020},
  publisher={IEEE}
}

@article{xu2020client,
  title={Client selection and bandwidth allocation in wireless federated learning networks: A long-term perspective},
  author={Xu, Jie and Wang, Heqiang},
  journal={IEEE Transactions on Wireless Communications},
  volume={20},
  number={2},
  pages={1188--1200},
  year={2020},
  publisher={IEEE}
}

@inproceedings{mcmahan2017communication,
  title={Communication-efficient learning of deep networks from decentralized data},
  author={McMahan, Brendan and Moore, Eider and Ramage, Daniel and Hampson, Seth and y Arcas, Blaise Aguera},
  booktitle={Proc. of International Conference on Artificial Intelligence and Statistics},
  pages={1273--1282},
  year={2017},
}

@article{xie2019asynchronous,
  title={Asynchronous federated optimization},
  author={Xie, Cong and Koyejo, Sanmi and Gupta, Indranil},
  journal={arXiv preprint arXiv:1903.03934},
  year={2019}
}

@inproceedings{elmahallawy2022asyncfleo,
  title={{A}sync{FLEO}: Asynchronous federated learning for {LEO} satellite constellations with high-altitude platforms},
  author={Elmahallawy, Mohamed and Luo, Tie},
  booktitle={2022 IEEE International Conference on Big Data (Big Data)},
  pages={5478--5487},
  year={2022},
  organization={IEEE}
}

@inproceedings{razmi2022scheduling,
  title={Scheduling for ground-assisted federated learning in {LEO} satellite constellations},
  author={Razmi, Nasrin and Matthiesen, Bho and Dekorsy, Armin and Popovski, Petar},
  booktitle={2022 30th European Signal Processing Conference (EUSIPCO)},
  pages={1102--1106},
  year={2022},
  organization={IEEE}
}

@article{elmahallawy2024communication,
  title={Communication-efficient federated learning for {LEO} constellations integrated with HAPs using hybrid NOMA-OFDM},
  author={Elmahallawy, Mohamed and Luo, Tie and Ramadan, Khaled},
  journal={IEEE Journal on Selected Areas in Communications},
  volume={42},
  number={5},
  pages={1097--1114},
  year={2024},
  publisher={IEEE}
}

@article{lin2024fedsn,
  title={Fed{SN}: A federated learning framework over heterogeneous {LEO} satellite networks},
  author={Lin, Zheng and Chen, Zhe and Fang, Zihan and Chen, Xianhao and Wang, Xiong and Gao, Yue},
  journal={IEEE Transactions on Mobile Computing},
  year={2025},
  volume={24},
  number={3},
  pages={1293-1307},
  publisher={IEEE}
}

@article{li2020federated,
  title={Federated learning: Challenges, methods, and future directions},
  author={Li, Tian and Sahu, Anit Kumar and Talwalkar, Ameet and Smith, Virginia},
  journal={IEEE signal processing magazine},
  volume={37},
  number={3},
  pages={50--60},
  year={2020},
  publisher={IEEE}
}

@inproceedings{wei2020efficient,
  title={An efficient {LEO} global navigation constellation design based on Walker constellation},
  author={Wei, Yanbo and Li, Huaijian and Du, Xiaojing},
  booktitle={2020 IEEE Computing, Communications and IoT Applications (ComComAp)},
  pages={1--6},
  year={2020},
  organization={IEEE}
}

@article{mo2016multi,
  title={Multi-Objective Optimization Design of {LEO} Satellite Constellations for Communication [D]},
  author={Mo, Y},
  journal={National University of Defense Technology},
  year={2016}
}

@article{kokkoniemi2021channel,
  title={Channel modeling and performance analysis of airplane-satellite terahertz band communications},
  author={Kokkoniemi, Joonas and Jornet, Josep M and Petrov, Vitaly and Koucheryavy, Yevgeni and Juntti, Markku},
  journal={IEEE Transactions on Vehicular Technology},
  volume={70},
  number={3},
  pages={2047--2061},
  year={2021},
  publisher={IEEE}
}

@inproceedings{xie2024mh,
  title={{MH}-p{FLID}: model heterogeneous personalized federated learning via injection and distillation for medical data analysis},
  author={Xie, Luyuan and Lin, Manqing and Luan, Tianyu and Li, Cong and Fang, Yuejian and Shen, Qingni and Wu, Zhonghai},
  booktitle={Proc. of International Conference on Machine Learning (ICML)},
  pages={54561--54575},
  year={2024}
}

@inproceedings{li2025mergenet,
  title={Mergenet: Knowledge migration across heterogeneous models, tasks, and modalities},
  author={Li, Kunxi and Zhan, Tianyu and Fu, Kairui and Zhang, Shengyu and Kuang, Kun and Li, Jiwei and Zhao, Zhou and Wu, Fan and Wu, Fei},
  booktitle={Proc of the Association for the Advancement of Artificial Intelligence Conference (AAAI)},
  volume={39},
  number={5},
  pages={4824--4832},
  year={2025}
}

@INPROCEEDINGS{Helber2018Introducing,
  author={Helber, Patrick and Bischke, Benjamin and Dengel, Andreas and Borth, Damian},
  booktitle={Proc of IEEE International Geoscience and Remote Sensing Symposium (IGARSS)}, 
  title={Introducing Eurosat: A Novel Dataset and Deep Learning Benchmark for Land Use and Land Cover Classification}, 
  year={2018},
  volume={},
  number={},
  pages={204-207},
}

@INPROCEEDINGS{Masihi2025Terahertz,
  author={Masihi, Ahmad and Testolina, Paolo and Jornet, Josep M.},
  booktitle={Proc of IEEE International Conference on Communications Workshops (ICC Workshops)}, 
  title={Terahertz vs. {O}ptical for Inter-Satellite Links: A Comparative Analysis of Pointing Errors and System Performance}, 
  year={2025},
  volume={},
  number={},
  pages={964-970},
}

@INPROCEEDINGS{Aliaga2025Analysis,
  author={Aliaga, Sergi and Petrov, Vitaly and Singh, Tejinder and Alavirad, Mohammad and Repeta, Morris and Healy, Michael and Jornet, Josep M.},
  booktitle={Proc of IEEE Consumer Communications and Networking Conference (CCNC)}, 
  title={Analysis of Scintillation Effects in Terahertz Band Satellite Communications for 6{G} and Beyond}, 
  year={2025},
  volume={},
  number={},
  pages={1-6},
}

@article{hinton2015distilling,
  title={Distilling the knowledge in a neural network},
  author={Hinton, Geoffrey and Vinyals, Oriol and Dean, Jeff},
  journal={arXiv preprint arXiv:1503.02531},
  year={2015}
}

@inproceedings{beyer2022knowledge,
  title={Knowledge distillation: A good teacher is patient and consistent},
  author={Beyer, Lucas and Zhai, Xiaohua and Royer, Am{\'e}lie and Markeeva, Larisa and Anil, Rohan and Kolesnikov, Alexander},
  booktitle={Proc of the IEEE/CVF conference on computer vision and pattern recognition (CVPR)},
  pages={10925--10934},
  year={2022}
}

@techreport{fcc2025spectrum,
  author      = {{Federal Communications Commission (FCC)}},
  title       = {{FCC} Looks to Unleash More Spectrum for Satellite Spectrum Abundance},
  institution = {FCC},
  year        = {2025},
  month       = {May},
  number      = {FCC-25-29},
  type        = {Notice of Proposed Rulemaking},
  url         = {https://www.fcc.gov/document/fcc-looks-unleash-more-spectrum-satellite-spectrum-abundance}
}

\vfill
  
\end{document}